\documentclass[journal=aamick,manuscript=article]{achemso}

\usepackage[version=3]{mhchem} % Formula subscripts using \ce{}
\usepackage{xcolor}
\usepackage{booktabs}
\usepackage{soul}
\usepackage{enumitem}
\usepackage{hyperref}
\author{Ashlee M. Garc\'ia}
\email{ashleeg@stanford.edu}
\affiliation[Stanford]
{Stanford University, 476 Lomita Mall, McCullough Bldg,  Stanford, CA 94305, USA}
\author{Hosni A. Kaissi}
\affiliation[UMinn]{University of Minnesota, 115 Union street SE, 
140 Physics \& Nanotechnology Building,
Minneapolis, MN 55455, USA}
\author{Jarod E. Meyer}
\author{Kira J. Martin}

\affiliation[Stanford]
{Stanford University, 476 Lomita Mall, McCullough Bldg,  Stanford, CA 94305, USA}

\author{Maksim Gomanko}
\affiliation[UPitt]{University of Pittsburgh, 3941 O’Hara St, 100 Allen Hall, Pittsburgh, PA 15213, USA}

\author{Laura A. Stern}
\author{Pooja D. Reddy}
\author{SeongJin Park}
\affiliation[Stanford]
{Stanford University, 476 Lomita Mall, McCullough Bldg,  Stanford, CA 94305, USA}
\author{Wilson J. Y\'anez-Parre\~no}
\affiliation[UMinn]{University of Minnesota, 115 Union street SE, 
140 Physics \& Nanotechnology Building,
Minneapolis, MN 55455, USA}
\author{Sergey M. Frolov}
\affiliation[UPitt]{University of Pittsburgh, 3941 O’Hara St, 100 Allen Hall, Pittsburgh, PA 15213, USA}
\author{Vlad S. Pribiag}
\affiliation[UMinn]{University of Minnesota, 115 Union street SE, 
140 Physics \& Nanotechnology Building,
Minneapolis, MN 55455, USA}

\author{Kunal Mukherjee}
\email{kunalm@stanford.edu}
\affiliation[Stanford]
{Stanford University, 476 Lomita Mall, McCullough Bldg,  Stanford, CA 94305, USA}

\title[]
  {Highly lattice-mismatched selective area epitaxy and coalescence of PbSe nanostructures on GaAs}

\begin{document}


\begin{abstract}

Selective area growth of PbSe has the potential to realize deterministic placement of high-density, defect-tolerant nanostructure networks toward a scalable quantum platform. PbSe is a narrow bandgap semiconductor with advantageous (opto)electronic properties that has been shown to have a desirable defect-tolerance, enabling bright emission even when grown on highly-dissimilar platforms, and could be leveraged in combination with selective growth for site-selective quantum emitters and low-disorder hybrid nanowire networks. In this work, we achieve site-selective growth of well-faceted and ordered PbSe nanostructures, despite a large 8\% lattice mismatch. Structural and morphological characterization reveal that $>$99\% of selectively grown islands within 100 nm opening are single-orientation and cube-on-cube oriented with sub-nm root-mean-square surface roughness. \textcolor{black}{Defect analysis showed that 74\% of the islands and 86\% of the coalesced regions between different mask openings were free of threading dislocations.} These PbSe islands achieved equivalent emission in the mid-infrared despite having a higher surface-to-volume ratio than the planar control. \textcolor{black}{Further, we present a gate-tunable two-terminal Josephson junction fabricated from the PbSe nanowires grown with conditions identified in this study. The combination of the accessible selective growth regime, morphological control of island growth and demonstrations of optical and electrical transport properties indicates promise for PbSe SAG as a defect-tolerant scalable quantum platform.}
\end{abstract}

\section{Introduction}
Selective area growth (SAG) by molecular beam epitaxy (MBE) of PbSe and other IV-VI chalcogenides offers an etch-free route for integrated photonics and quantum technologies. SAG has re-emerged as a critical planar processing technique for realizing direct epitaxial growth toward \textcolor{black}{achieving} high-fidelity, scalable physical qubits\cite{Beznasyuk2022,friedl_template-assisted_2018, lee_selective-area_2019, seidl_postgrowth_2021, krizek_field_2018,jung_selective_2022} %hertel_gate-tunable_2022, hertel_electrical_2021, vaitiekenas_selective-area-grown_2018,} 
and all-epitaxial, laterally structured photonics.\cite{Kafi,Skipper2022,garcia_high-quality_2025,Ironside2019} By using an amorphous mask to define crystal growth, geometric control and deterministic placement of epitaxially smooth nanostructures can be achieved without post-growth patterning techniques, which introduce disorder and degrade quantum efficiency. Further, subsequent epitaxial lateral overgrowth and planar coalescence realizes the seamless integration of \textcolor{black}{metals and dielectrics} into crystalline semiconductor,\cite{Ironside2019,garcia_high-quality_2025} which could enable advanced photonic and electronic device designs with embedded photonics, metasurfaces, and contacts in combination with IV-VI chalcogenides.

IV-VI rocksalt chalcogenide semiconductors are attractive for SAG because their narrow direct bandgaps,\cite{ekuma_optical_2012,suzuki_optical_1995} strong spin orbit coupling,\cite{khan_thermoelectric_2016} low Auger recombination rates for minority carriers,\cite{klann_fast_1995,boberl_midinfrared_2003} and high dielectric constants\cite{allgaier_mobility_1958,springholz_wiley_2014} may enable a disorder resilient functionality for mid-infrared photonics and quantum devices, including but not limited to band-engineered nanostructures for spin and superconducting qubits, topological crystalline insulator platforms and near- to long-wave infrared tunable quantum emitters.
When integrated on highly dissimilar platforms, IV-VI semiconductors have the potential to outperform incumbent III-Vs due to this inherent defect tolerance. PbTe SAG on InP templates recently established this semiconductor for hybrid quantum nanowire devices,\cite{jung_selective_2022} exploring the synergistic pairing of these intrinsic transport properties with the etch-free SAG technique.  Those in-plane nanowire networks showed Hall bar mobilities up to 5600 $cm^2(Vs)^{-1}$ and a low-temperature Aharonov-Bohm phase coherence exceeding 21 $\mu$m despite a large $\sim$10\% lattice mismatch between the film and substrate, demonstrating its promise as a material platform for on-chip quantum systems.\cite{jung_selective_2022} 
The ability to leverage the inherent advantageous properties of PbTe via SAG on a III-V platform motivates the exploration of this technique for other IV-VI chalcogenides, most notably its selenide counterpart. 

PbSe holds promise for optoelectronics and hybrid semiconductor-superconductor networks as it shares many advantageous properties with PbTe, but also has been shown to be a more efficient light emitter than PbTe due to a higher band edge density of states and lower nonradiative recombination rates.\cite{klann_fast_1995,boberl_midinfrared_2003} Prior PbSe-on-GaAs films\cite{meyer_bright_2021,haidet_interface_2021} and light-emitting diodes\cite{meyer_midaeinfrared_2025} retain strong emission despite dislocation densities exceeding 10$^9$ cm$^{-2}$. Thus, PbSe SAG on a technologically-relevant (001) III-V platform, in particular the cost-effective GaAs, has the potential to bring together deterministic nanostructure placement, coalescence-aware \textcolor{black}{dislocation} management, and defect-tolerant mid-infrared optical and hybrid quantum devices in a highly mismatched IV–VI/III–V system (Figure~\ref{fig:selectivity}b-d).  

However, the broader materials science question remains unresolved: how do \textcolor{black}{selectively grown} IV–VI islands nucleate, laterally overgrow, coalesce, and generate dislocations while mediating such large lattice mismatch? Specifically, the conditions necessary to achieve both deposition selectivity (i.e. preferred growth in mask openings) and growth of \textcolor{black}{high-quality} PbSe nanostructures with well-ordered coalescence need to be studied. Developing an understanding of how its rocksalt crystal structure drives the resulting growth geometry and morphology would enable the precise growth of PbSe nanostructures and additionally act as a scaffold for the selective growth of its cubic chalcogenide alloys (e.g. PbSnSe\cite{meyer_engineering_2024,reddy_expanded_2024,reddy_reversible_2025,dziawa_topological_2012,henini_chapter_2018}, PbGeSe\cite{xiao_epitaxial_2024,nikolic_solid_1969,krebs_uber_1964,luo_soft_2018}).

In this work, we demonstrate selective area epitaxy, facet-controlled overgrowth and partial coalescence of PbSe islands and nanowires on 8\% lattice-mismatched GaAs and highlight a path toward defect-tolerant integrated photonic and quantum platforms. First, \textcolor{black}{we identify a range of substrate temperatures where growth of PbSe is achieved within the mask openings and polycrystalline deposition on the mask is mitigated}. We then demonstrate smooth, ordered, and \textcolor{black}{well-faceted} growth at 365 $^\circ$C\textcolor{black}{; the morphology of these islands was observed to be driven by forming and maintaining low-energy \{001\} facets\cite{deringer_stabilities_2016} with sub-nanometer root-mean-square (RMS) roughnesses. The observed maintenance of the (001) surface contrasts other semiconductor systems (e.g. III-Vs\cite{Ironside2021,white_selective_2026,garcia_high-quality_2025,Ironside2019}) and is expected to be advantageous for subsequent planar coalescence over embedded dielectric structures. Further, islands grown in 100 nm wide square} openings were $>$99\% single-orientation, free of stacking faults, and \textcolor{black}{74\%} threading-dislocation-free despite the lattice mismatch and dissimilar crystal structures\textcolor{black}{, which contrasts prior demonstrations of site-selective nanoheteroepitaxy \cite{hausler_investigation_2024, langdo_high_2000, Li2007} and lateral epitaxial overgrowth\cite{langdo_high_2000,suwannaharn_structural_2022,knoedler_observation_2017,lee_selective-area_2019, sun_optical_2015}.} \textcolor{black}{From these islands, equivalent room-temperature emission in the mid-infrared was measured in comparison to a planar control, despite the lower material volume and increased surface-to-volume ratio, showing promising optical quality.} \textcolor{black}{Finally, gate-modulated supercurrents were demonstrated in a Josephson field effect transistor (Jo-FET) fabricated from the PbSe SAG nanowires with Al superconducting contacts.} \textcolor{black}{Overall, this work demonstrates the promise of PbSe SAG as a scalable platform for realizing quantum architectures.} 

\section{Results and Discussion}

\subsection{PbSe selective growth window}

\begin{figure}
    \centering
    \includegraphics[width=1\linewidth,clip,trim=0in 5.7in 0in 0in]{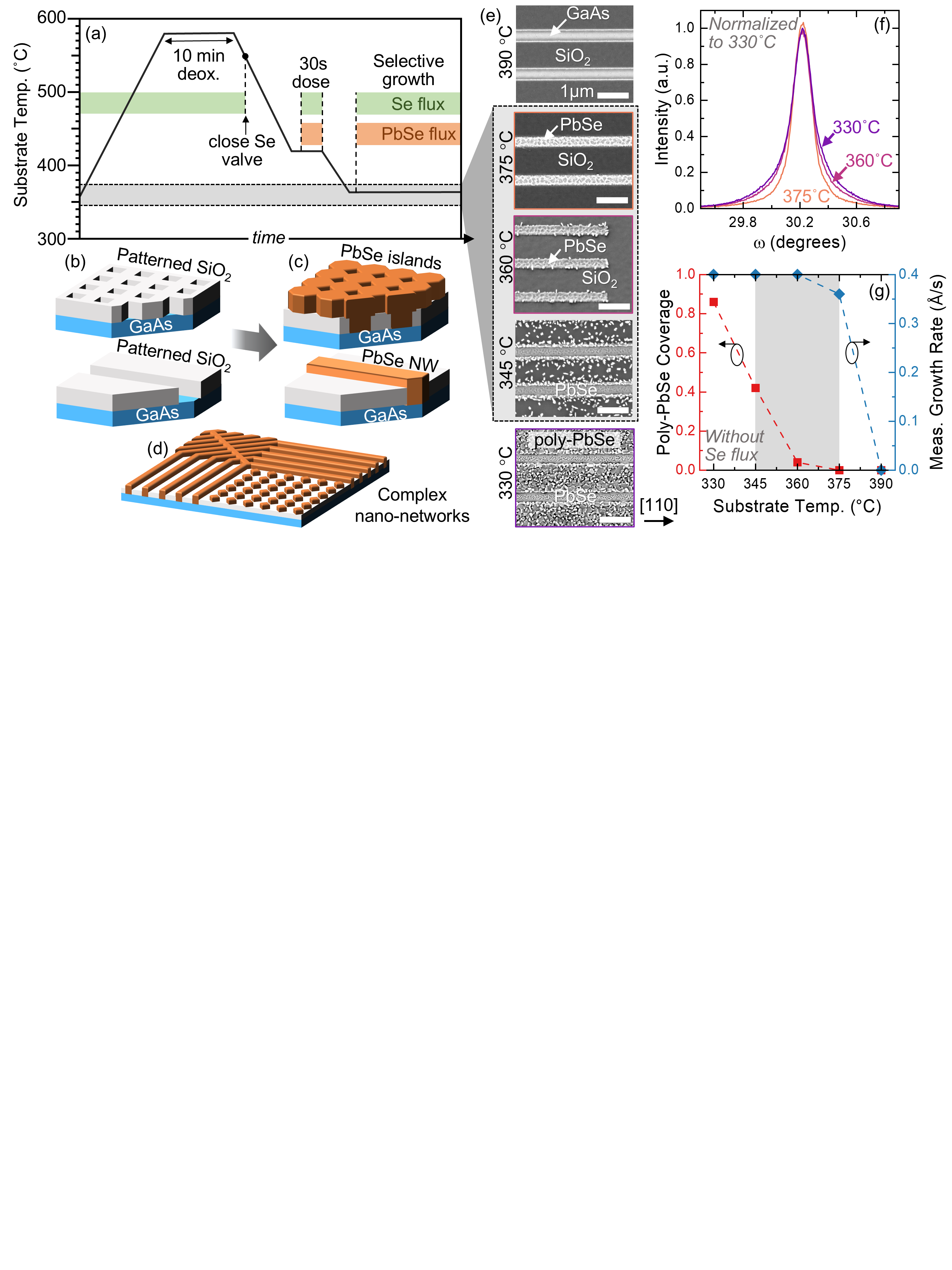}
    \caption{(a-d) Schematic representation of selective area growth (SAG) technique and (e-g) characterization of the selective growth window of PbSe over a patterned SiO\textsubscript{2} mask. (a) In-situ template preparation and growth conditions employed to achieve cube-on-cube growth confined to mask opening are represented in schematic graph of temperature versus time. Diagram representations of (b) initial lithographically-defined patterned mask template and (c) confined PbSe growth forming low-disorder nanostructures, both islands and nanowires (NWs), for scalable integrated photonics and quantum platforms based on (d) complex nano-networks. (e) Planview SEM images after 50 nm PbSe deposition on a 47 nm mask at substrate temperatures varying from 330 $^\circ$C to 390 $^\circ$C. Preferred growth of PbSe in the GaAs windows occurred between 345 $^\circ$C and 375 $^\circ$C. The scale bars are 1 $\mu$m. (f) Plot of (004) PbSe rocking curves of the planar control films grown simultaneously alongside the selective growth templates demonstrating no added broadening from use of higher substrate temperatures. (g) Graph of mechanisms bounding the selective growth window versus substrate temperature. Polycrystalline PbSe areal coverage of the SiO\textsubscript{2} is plotted on left axis and effective PbSe growth rate measured by the resulting thickness of these samples is plotted on the right axis. }
    \label{fig:selectivity}
    
\end{figure}

\textcolor{black}{Initial work in this study} aimed to establish a selective growth window for PbSe over patterned SiO\textsubscript{2} on (001) GaAs, leveraging the typical process flow (i.e. deoxidation, PbSe dose and subsequent growth\cite{haidet_interface_2021}) used to achieve cube-on-cube growth in planar films illustrated in Figure~\ref{fig:selectivity}a. The substrate temperature during growth was varied from the typical planar growth conditions, at 330 $^\circ$C, up to 390 $^\circ$C, where re-evaporation of material is significant\cite{springholz_molecular_2017} and no PbSe growth was observed in mask openings. Characteristics of the polycrystalline PbSe formation on the SiO\textsubscript{2} mask (e.g. surface coverage or fraction\cite{garcia_surface_2026}, depletion width\cite{Allegretti1995}) were used to \textcolor{black}{characterize the} selective growth regime.  As shown in Figure~\ref{fig:selectivity}e, near-complete polycrystalline PbSe coverage of the mask at 330 $^\circ$C indicated little desorption of PbSe from the mask surface during growth. As temperature increased to 345-360 $^\circ$C, preferred growth in openings of the mask \textcolor{black}{was} observed with $<50\%$ polycrystalline coverage on the mask surface. At 375 $^\circ$C thermal desorption of PbSe off of the mask surface was observed to be greater than its adsorption rate, enabling a complete suppression of polycrystalline deposition. These increases in growth temperature resulted in little variation in crystal quality as evidenced by the absence of additional broadening in the PbSe(004) rocking curves of the planar control films (Figure~\ref{fig:selectivity}f), indicating that the selective growth window remains compatible with typical PbSe epitaxy quality. However, the \textcolor{black}{thickness of the growth} at 375 $^\circ$C saw a \textcolor{black}{$\sim$10}\% \textcolor{black}{reduction compared to the unity-sticking growth rate as shown in Figure~\ref{fig:selectivity}g.} 
 
Therefore, subsequent studies used a compromise growth temperature of 365 $^\circ$C combined with excess Se fluxes corresponding to 3$\times10^{-8}$ and 3$\times10^{-9}$ Torr beam-equivalent-pressures (BEPs) during the growth and dose steps, respectively, to limit re-evaporation driven roughening. 
Here, deposition selectivity was observed to be dependent on diffusion of PbSe from the mask to the growth window. \textcolor{black}{The largest polycrystalline-free mask area was $\sim$200 nm wide, therefore the diffusion length of PbSe on the mask was approximated to be $\sim$100 nm,\cite{Allegretti1995} which is comparable to the observed low adatom diffusion lengths on the mask during III-V SAG demonstrations.\cite{Lee2016}}

\subsection{Nucleation control and faceted lateral epitaxial overgrowth}
\textcolor{black}{With the selective growth window established, we next evaluated whether the same conditions could also enable PbSe nucleation aligned to the underlying substrate and achieve morphologically-controlled lateral epitaxial overgrowth (LEO) above the mask surface.}

\begin{figure}
    \centering
    \includegraphics[scale=0.8,clip,trim=0in 3.02in 0.42in 0in]{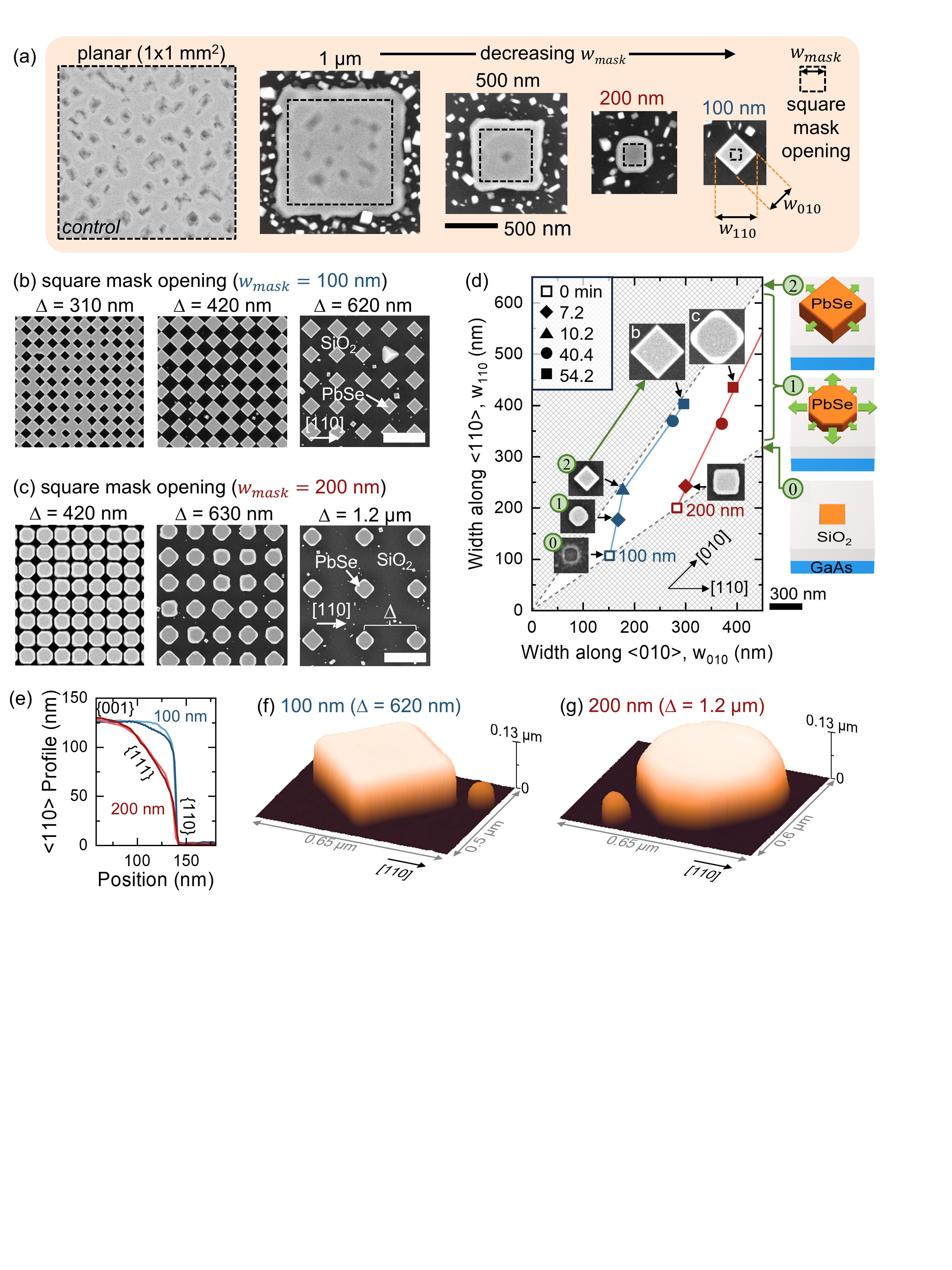}
    \caption{(a) SEM images of 180 nm PbSe grown within square mask openings of various widths ($w_{mask}$=1 $\mu$m, 500 nm, 200 nm, 100 nm) patterned in an 83 nm tall SiO\textsubscript{2} film compared to the planar control region, highlighting the improvement in surface morphology observed for  $w_{mask}\leq$ 200 nm. (b-c) Planview SEM images show growth morphology of (b) 100 nm and (c) 200 nm wide squares spaced periodically by specified lengths $\Delta$, after 180 nm of PbSe growth at 365 $^\circ$C over a 50 nm tall SiO\textsubscript{2} mask. Scale bar is 1 $\mu$m. (d) Graph showing the evolution of the PbSe growth shape, characterized by its widths along the $<$010$>$ and $<$110$>$ directions, with respect to different overgrowth times, where t=0 min is when PbSe growth is in-line with the top of the SiO\textsubscript{2} mask. \textcolor{black}{Grey dotted lines represent the geometric relationship between $w_{110}$ and $w_{010}$ that bound the initial mask square ($w_{010}=w_{110}\sqrt2$) and final LEO geometry ($w_{110}=w_{010}\sqrt2$).} Inset SEM images are representative of the average PbSe shape at each their respective time steps showing the two stages of growth and schematics to the right of the graph indicate the preferred growth directions at each stage. (e) Graph of the [110] and [-110] profiles at the widest point of the PbSe overgrowth from the AFM scans of (f) 100 nm ($\Delta$=620 nm) and (g) 200 nm ($\Delta$=1.2$\mu$m) regrown squares from (b) and (c), respectively.}
    \label{fig:squares}
    
\end{figure}

\textcolor{black}{Under the optimized 365 $^\circ$C conditions, we found that growth in square mask openings less than 200 nm wide produced visually smooth and faceted islands after 180 nm of PbSe growth over an 83 nm thick SiO\textsubscript{2} mask (Figure~\ref{fig:squares}a). This study was then repeated over a mask template with a 50 nm thick SiO\textsubscript{2} layer and it showed that these PbSe SAG conditions reliably produced near-deterministic cube-on-cube nucleation, realizing uniform arrays of visually well-faceted and ordered features (Figure~\ref{fig:squares}b,c).} We find that low-energy \{001\} facets dominate growth morphology,\cite{deringer_stabilities_2016} which is consistent with PbTe SAG on both (111) and (100) InP\cite{jung_selective_2022}. \textcolor{black}{Even though the square openings ($w_{mask}\leq200$ nm) in the SiO\textsubscript{2} were aligned to the $<$110$>$ cleave planes of the underlying GaAs wafer, the PbSe lateral overgrowth shape of the (001)-oriented nuclei \textcolor{black}{above the mask} appeared as larger squares rotated 45$^\circ$, with (100) and (010) sidewalls.} Due to the sensitivity of this heteroepitaxial growth process, it was observed that the SiO\textsubscript{2}/GaAs template preparation had considerable influence on the heteroepitaxial nucleation, details of which can be found in Supplemental Information. Optimized templates produced cube-on-cube oriented nuclei  within 99.7\% and 100\% of the 100 nm and 200 nm square mask openings, respectively.

Developing a comprehensive understanding of how the confined epitaxial interface evolves with continued material deposition, particularly as growth proceeds over the mask, is essential for achieving morphology-controlled nanostructures\cite{cachaza_selective_2021,dede_selective_2022} and/or high-quality planar coalescence over embedded mask.\cite{Ironside2019,Ironside2021,garcia_high-quality_2025,raftery_buried_2026,raftery_photopumped_2025} The primary determinants of the lateral epitaxial overgrowth (LEO) morphology are anticipated to be the orientation and geometric shape of the mask openings, in conjunction with the competition between energetically favorable crystallographic planes.\cite{takebe_orientation-dependent_1997,nishinaga_surface_1996} A time-dependent LEO study revealed a two-stage faceting process that converts the original $<$110$>$-aligned mask geometry into a stable, diagonally oriented square bounded by \{001\}-type sidewalls. The average lateral dimensions along the $<$110$>$ and $<$010$>$ were measured at different overgrowth heights for 100 nm and 200 nm wide square openings (Figure~\ref{fig:squares}d). In this study, $t=0$ corresponds to when the lateral dimensions of PbSe are bound by the mask opening (i.e the PbSe growth height $\leq$ SiO\textsubscript{2} mask height). 
As the PbSe begins to grow above the SiO\textsubscript{2} surface, the $<$110$>$ directions saw the fastest lateral growth rate. This defines the initial stage of LEO where the in-plane shape transitions from a $<$110$>$-aligned square to an octagon; the evolution is driven by preference for incorporation on the \{110\} planes \textcolor{black}{at average lateral growth rates of 0.5 and 0.3 \AA/s for 100 nm and 200 nm islands, respectively}, while growth along the $<$010$>$ direction \textcolor{black}{proceeds} at a slower average rate of $\sim$0.2 \AA/s. The second stage occurs once the $\{$110$\}$ plane of this fast-growing direction is eclipsed, forming \textcolor{black}{the diagonally-oriented square geometry} with its sides defined by the $<$010$>$ in-plane directions as opposed to the original $<$110$>$ orientation. In this stage, there is a redistribution of the diffusive contribution to all \{001\} surfaces and this shape is maintained.

Unlike the growth in the smaller $\sim$100 nm square mask openings, PbSe islands grown in the larger $\sim$200 nm mask openings remained octagonal due to the larger overgrowth volume needed to fully \textcolor{black}{eclipse} the \{110\} planes. The slower growth rate in this direction for larger features allowed for competition with other slower forming crystal planes. Comparing the height profiles along [110] attained via atomic force microscopy (AFM), it is evident that \{111\} planes formed on the $\sim$200 nm features as shown by a 54.7$^\circ$-angled section of the sidewall (Figure~\ref{fig:squares}e-g). The formation of this plane \textcolor{black}{is consistent} with the comparable surface energies of the \{110\} and \{111\} planes\cite{deringer_stabilities_2016} and matches the observed facet preference for PbTe SAG in 200 nm (diam.) circular openings on InP(100)\cite{jung_selective_2022}. \textcolor{black}{The formation of the \{111\} planes along dimensions $\geq$200 nm while maintaining a preference for the flat \{001\} top surface is promising for coalescence of PbSe over periodic-grating structures using a two-step embedded regrowth approach.\cite{Ironside2019} It would enable the fast-growing lateral fronts to meet at a one-dimensional line while minimizing the amount of material (i.e. typically few microns for III-Vs\cite{garcia_high-quality_2025,Ironside2019} and III-Ns\cite{li_gan_1999,suwannaharn_structural_2022}) needed to fully restore a smooth (001) surface above micron-wide gratings via planar coalescence.}

The selective growth geometry confined to below 200 nm openings was found to improve surface morphology relative to planar PbSe. AFM measurements confirmed smooth top \{001\} facets of the PbSe islands. The average RMS roughness of the individual islands in the 100 nm and 200 nm mask openings in Figure~\ref{fig:squares}f,g was 0.5 and 0.9 nm, respectively, indicating that growth at 365 $^\circ$C did not suffer from entropic roughening, which is common in demonstrations of III-V SAG.\cite{Ironside2019,Fahed2016-uj} Further, the smooth surfaces appeared to be a benefit of the selective growth technique confining nucleation and initial coalescence of grains to a small area. As shown in Figure~\ref{fig:squares}a, when the square mask openings increased past 200 nm in width to 500 nm and 1 $\mu m$, non-uniform surfaces consistent with the morphology of the planar region (RMS $\sim3.6$ nm) were observed. The non-uniform morphology results from the coalescence of multiple faceted PbSe grains with pits at their intersection. A visually smooth border, approximately 270 nm in width, combined with the pitted center region for 1 $\mu m$ and 500 nm wide openings, is indicative of a potential growth rate enhancement\cite{li_gan_1999} from diffusion off the SiO\textsubscript{2} mask to the growth window that achieves better intra-island PbSe uniformity in the square openings $\leq$ 200 nm.

We identify inter-island height variability as the remaining morphology limitation, particularly in the 100 nm openings. Despite the smooth \{001\} surfaces, there was significant height variation between islands grown in the 100 nm square mask opening arrays; the island heights had an average range (e.g. Z=max$-$min) of 18.2 nm, similar to that of growth in the planar region (Z=23.5 nm). We hypothesize that this may be due to time variability in the initial nucleation between sites, which becomes exacerbated with continued growth. Interestingly, \textcolor{black}{the range of island heights from growth in the 200 nm squares} was only 2.6 nm, $7\times$ more narrow than the islands in 100 nm squares, giving further confidence to this hypothesis. The initial nucleation event that dominates the growth orientation of PbSe islands in 200 nm squares likely occurs in a narrower time frame because its larger area accommodates the low initial nucleation density. Increasing the initial nucleation density of PbSe on GaAs via optimization of the surface pre-treatment dose or leveraging a colder initial nucleation step could mitigate the variability observed in smaller mask openings. In summary, PbSe SAG combines near-unity cube-on-cube nucleation, smooth \{001\} top facets, predictable lateral faceting, and improved morphology relative to planar growth. These features establish the structural preconditions for controlled coalescence and defect reduction, discussed next.

\begin{figure}
    \centering

    \includegraphics[width=1\linewidth,clip,trim=0in 4.5in .6in 0in]{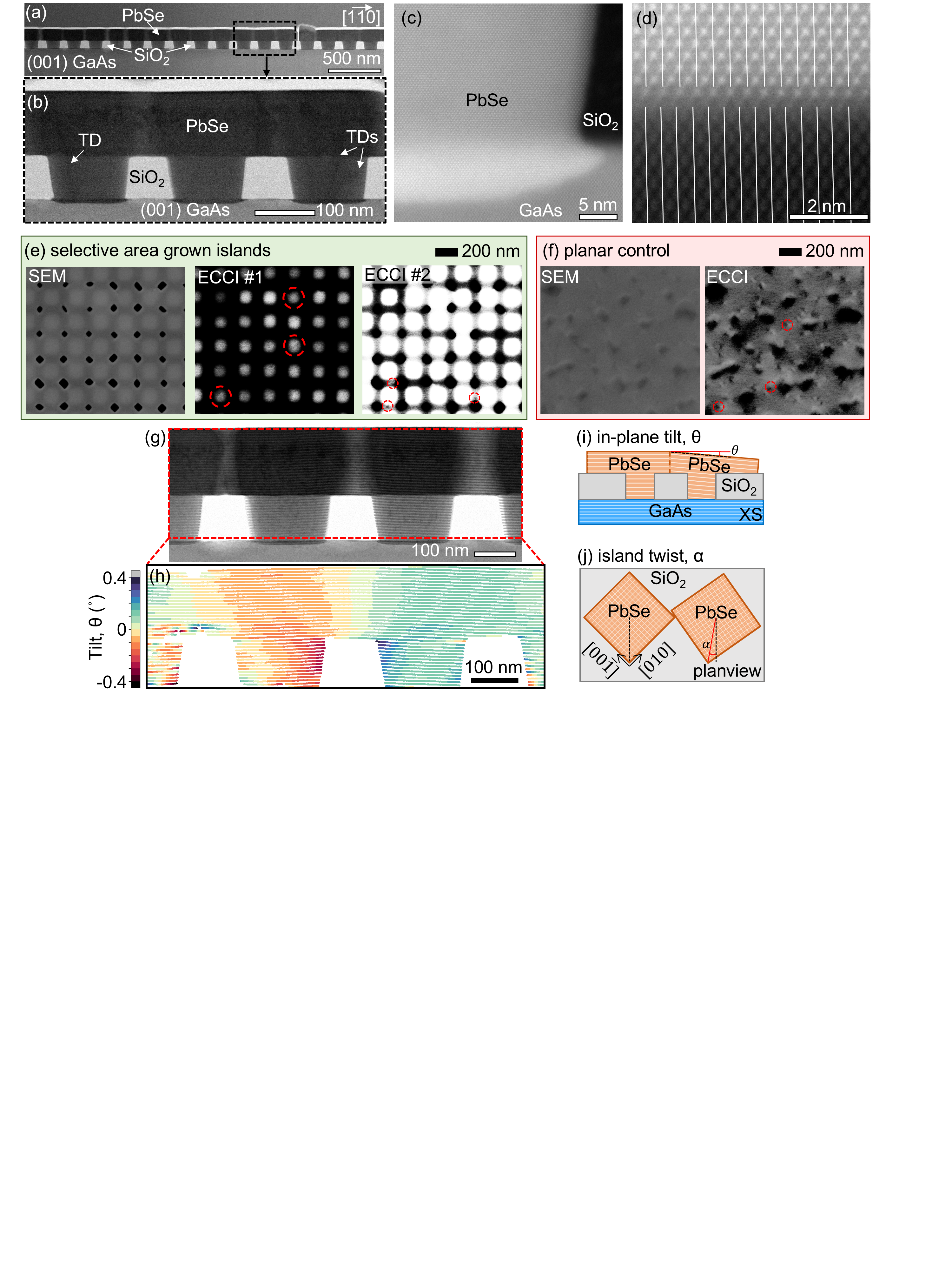}
    \caption{(a) \textcolor{black}{Bright-field} (BF) STEM cross-section of 180 nm PbSe SAG over an 83 nm tall mask patterned with 100 nm wide square openings \textcolor{black}{in addition to a (b) higher-resolution cross-section.} \textcolor{black}{High-resolution images of the (c) oxide/PbSe/GaAs interface and (d) the nucleation interface of PbSe on (001) GaAs. (e) Electron channeling contrast imaging of PbSe islands. From left to right, SEM image, low contrast ECCI to visualize TDs formed at the PbSe/GaAs interface and high contrast conditions to visualize threading dislocations from coalescence of laterally grown facets. (f) SEM image and corresponding ECCI of planar grown PbSe control region. \textcolor{black}{(e,f) For illustrative purposes, dashed circles highlight three TDs in each ECCI measurement.} (g) Cross-sectional BF STEM image showing a moir\'e fringe pattern and (h) corresponding colormap of measured tilt of PbSe planes and illustrating the merging of islands. (i) Planview and (j) cross-section schematics display the twist, $\alpha$, and tilt, $\theta$, of the PbSe grains with respect to the orientation of the underlying wafer.}}
    \label{fig:TEM}
\end{figure}

\subsection{Mismatch accommodation and coalescence-driven defect formation}

The 8\% PbSe/GaAs lattice-mismatch and dissimilar crystal structures raise a central question for SAG: How do isolated islands and partially coalesced films accommodate mismatch, and do they still generate the high threading-dislocation densities typical of blanket heteroepitaxy?    
To examine \textcolor{black}{island} microstructure and dislocation content, a thick 180 nm growth over an 83 nm SiO\textsubscript{2} tall mask (t=40.4 min in Figure~\ref{fig:squares}c), which exhibited partially-coalesced islands, was characterized by scanning transmission electron microscopy (STEM) and electron channeling contrast imaging (ECCI)\cite{kamaladasa_basic_2010}. Cross-sectional STEM images were prepared along the [110] zone axis, capturing several island/GaAs interfaces as well as coalescence regions (Figure~\ref{fig:TEM}). We find that PbSe SAG on GaAs is uniform and well-behaved morphologically, accounting for foil thickness due to the geometry of the islands and coalescence air-gaps. A magnified image of the interface (Figure~\ref{fig:TEM}b) shows some undercut of the GaAs from dielectric etching, and some subsurface damage via the strain contrast.  Nevertheless, the PbSe layer was largely able to accommodate this subsurface damage as well as the 8\% lattice mismatch and achieve cube-on-cube orientations without affecting crystal quality. The heterointerface appeared incoherent and yet sharp (i.e. the lattice of PbSe smoothly disregisters across the lattice of GaAs without localized misfit dislocations), which is in line with prior work\cite{haidet_nucleation_2020} on blanket growth of PbSe on GaAs on higher quality As-capped GaAs templates. We highlight some threading dislocations within islands in the STEM image (Figure~\ref{fig:TEM}b), however, a majority of the interior of the islands themselves are threading dislocation-free. We propose that mismatch relaxation occurs directly at the incoherent interface, uniquely circumventing the need to nucleate threading and misfit dislocations. This is reminiscent of the highly mismatched GaSb-on-GaAs films that also show unconventional mismatch relief, with the notable difference being semi-coherent interfaces in the latter as an interfacial misfit dislocation array is present. \cite{qian_nucleation_1997,huang_strain_2006,huang_interfacial_2009} 

ECCI shows with more certainty that selective growth reduces the threading dislocation density (TDD) relative to blanket PbSe, leaving most 100 nm islands threading-dislocation-free. We used ECCI to highlight \textcolor{black}{threading dislocations (TDs)} inside the individual islands as well as in coalescence boundaries across different openings (Figure~\ref{fig:TEM}e) by varying image contrast. The islands grown in 100 nm square openings have an effective TDD of $2\times 10^9$ $cm^{-2}$, in contrast to $7\times 10^9$ $cm^{-2}$ measured in a blanket region (Figure~\ref{fig:TEM}f). Furthermore, 74\% of the islands for 100 nm openings were threading-dislocation-free. Interestingly, while we noted several threading dislocations in the coalescence boundaries of these islands, there were also many that appeared to merge without generating a defect. The net dislocation density of the film (island + coalescence) is a factor of 3.5 lower than the blanket film for 100 nm openings. Furthermore, we did not observe stacking faults, which have traditionally been a challenge for selective heteroepitaxy and epitaxial lateral overgrowth, particularly along the [110] direction, of other materials, such as the III-V\cite{sun_optical_2015,lee_selective-area_2019,knoedler_observation_2017} and group IV\cite{langdo_high_2000} systems. 

The observed threading dislocations can arise from three distinct pathways: (1) angular misorientation, i.e. tilts or twists, between neighboring islands that give rise to geometrically necessary threading dislocations upon coalescence, (2) lateral translational disregistry between the islands due to lattice-mismatch that also \textcolor{black}{results} in misfit and threading dislocations upon coalescence, and (3) multi-nucleation within a single mask opening uncorrelated in orientation and translation; essentially a combination of the first two mechanisms but occurring within an island.\cite{mcmahon_perspective_2018} Island coalescence-related TDs  ($2.1 \times10^9 cm^{-2}$) are most likely a result of island orientation mismatch due to the mosaic nature of PbSe nucleation. The number of geometrically necessary dislocations scales as $w_{coal}\alpha/b$ for twist and $h\theta/b$ for tilt, where $w_{coal}$ and h are the contact width and height at coalescence, $\alpha$ and $\theta$ are the respective misorientation twist and tilt angles, and $b$ is the Burgers vector magnitude. This translates to about 1 threading dislocation for a contact dimension of 60 nm with a misorientation angle of 0.4$^\circ$. In many instances, the island tilt or twist angles are small with respect to the critical contact dimensions and the strain is accommodated elastically without dislocations. At a select magnification in STEM, horizontal scanning moir\'e lines were visible in the PbSe layer (Figure~\ref{fig:TEM}g,h); these give a magnified view of localized lattice tilting in the islands. Tilting of the PbSe islands is visualized by a diagram in Figure~\ref{fig:TEM}i. The moir\'e fringes have varying tilt angles below the dielectric for each island, indicating mosaicity. As the islands grow above the dielectric mask and coalesce, the fringes smoothly transition and bend to accommodate these tilt variations and result in a high-quality merge with tilting accommodated elastically. From the fringe spacing and fringe tilt, we estimate a median lattice tilt difference of 0.4$^\circ$ and 0.2$^\circ$ for two sets of two-island merges studied, at or below the critical height criterion. We do not have twist angle mismatch as these require planview images, but the \textcolor{black}{TD density} suggests they must be close to the critical width threshold, occasionally yielding dislocations. 

We propose that the second mechanism of dislocations due to translational disregistry between islands is deferred due to the partially coalesced nature of the film. The open voids provide a topological escape mechanism for translational disregistry as the free surfaces can close Burgers circuits without a threading dislocation.\cite{mcmahon_perspective_2018} This makes the coalescence geometry central to defect control in large-mismatch SAG. The tendency of PbSe islands to develop $<$100$>$ facets, combined with the square mask geometry, produces corner-to-corner contact at a single point rather than edge-to-edge contact. This one-zipper geometry is known to be favorable for TD-free coalescence relative to two-zipper edge merges.\cite{yan_coalescence_2000} Thus, we expect that complete coalescence of these samples would result in new threading dislocations to appear at the locations of the voids. Further optimization of the mask design could reduce or reposition dislocations via: (1) greater extent of lateral overgrowth before coalescence, (2) improved etch profile to reduce damage and improve mosaicity, and (3) harness the natural facet geometry to selectively position coalescence dislocations. 

The intra-island TDs, with an effective density of $2.8\times10^9 cm^{-2}$, likely arise from the third mechanism of multi-nucleation within a single mask opening and early coalescence. This dislocation density is also amenable to engineering through nucleation control via moving to smaller mask openings, fabrication optimization, and changing growth conditions. In contrast, the 100 nm wide nanowire mask openings were subject to the merging of multiple nuclei and exhibited a threading dislocation density much closer to the blanket (Figure S9, Supplemental Information). In summary, we decomposed the dislocation content of the PbSe SAG samples across three controllable mechanisms, and showed that SAG produces ordered, faceted, selectively placed PbSe nanostructures with reduced net dislocation density relative to blanket growth. We next examine whether these nanostructures remain compatible with optically active and hybrid superconducting functionality that motivates this platform.

\begin{figure}
    \centering
    \includegraphics[page=4,scale=.78,clip,trim=0in 1.84in 4.04in 0in]{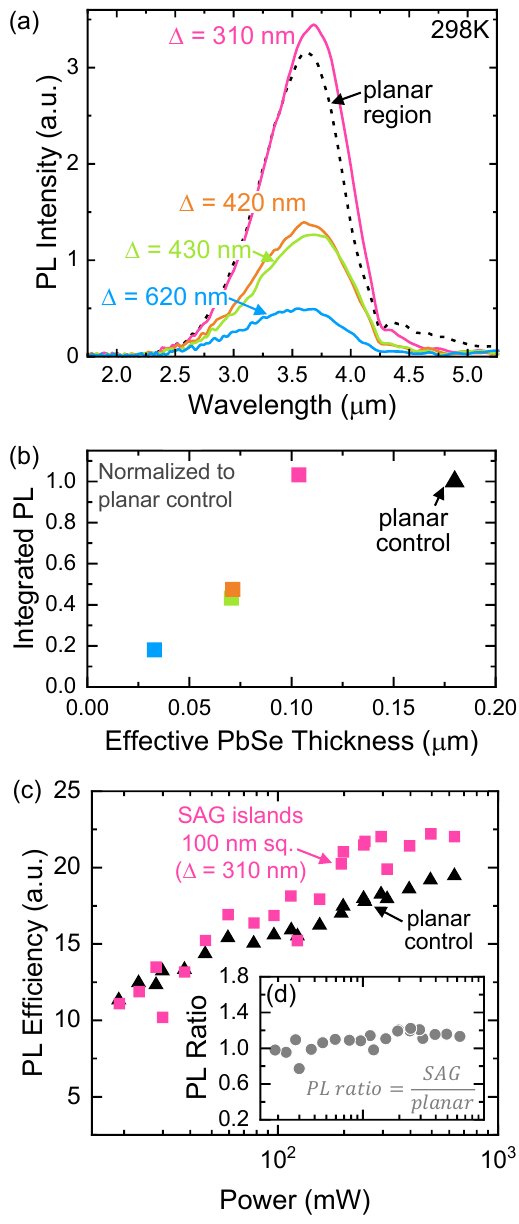}
    \caption{(a) Room temperature photoluminescence (PL) spectra of 100 nm square arrays, spaced periodically by pitch $\Delta$, regrown with 180 nm PbSe over a 50 nm tall SiO\textsubscript{2} mask compared to the emission from an unpatterned region of the substrate. (b) Graph of integrated PL intensity with respect to the effective film thickness (\textcolor{black}{i.e. PbSe SAG island volume averaged over area of array period to account for variable island width}) shows expected scaling of signal with PbSe SAG volume. Equivalent emission was observed for the highest density array compared to the planar control indicating no degradation from the SAG integration technique and potentially enhanced light absorption or extraction. (c) Plot of the PL efficiencies of the highest density SAG array (100 nm, $\Delta$=310 nm) and the planar control graphed with respect to pump power. (d) Inset graph of ratio between measured PL signals (SAG/planar) versus pump power, showing negligible variation in non-radiative mechanisms. }
    \label{fig:PLquality}
    
\end{figure}

\subsection{Room temperature mid-infrared emission from SAG PbSe arrays}
We show that selectively grown PbSe nanostructures retain strong room-temperature mid-infrared emission despite their reduced volume and increased surface area. Selectively grown PbSe islands were studied through photoluminescence (PL) spectroscopy.  Due to the small total volume of the island arrays, optical characterization was performed on the 100 nm wide square arrays (50 nm SiO\textsubscript{2} mask) that were regrown with 180 nm PbSe, which was the sample with the longest overgrowth time and therefore largest PbSe volume, highlighted in Figure~\ref{fig:squares}a. Room-temperature PL was measured using quasi-continuous-wave 808 nm excitation and FTIR detection, comparing SAG arrays to an unpatterned planar PbSe control region on the same substrate.  

Figure~\ref{fig:PLquality}a displays the PL spectra of the regrown regions compared to the planar control region and Figure~\ref{fig:PLquality}b shows the relationship between PL intensity and the volume of PbSe represented as effective film thickness (e.g. PbSe SAG volume divided by the area of the array period). The regrown regions showed room-temperature photoluminescence in the mid-infrared comparable to the planar region of the substrate in both the \textcolor{black}{emission wavelength} and signal magnitude, indicating that the selective growth process does not strongly perturb the PbSe band-edge emission. It was observed that the intensity of the PL signal from the regrown regions scales with the density of the array of square openings in the mask as anticipated due to the change in volume. The highest density array ($\Delta=310 nm$ in Figure~\ref{fig:squares}a) produced a PL signal equivalent to that of the planar control region despite the lower volume of PbSe and having substantially increased the surface area-to-volume ratio. 

The unexpectedly strong PL from the highest density SAG array likely reflects a combination of retained material quality and geometry-dependent enhanced absorption at the pump wavelength and increased emitted light extraction. PbSe has a high refractive index compared to air at both the pump ($n_{PbSe}\sim 5.2$)\cite{suzuki_optical_1995} and emission wavelengths ($n_{PbSe}\sim 4.8-5.0$)\cite{zemel_electrical_1965}, therefore sees significant reflection at the PbSe-air interface in planar films which affects light absorption and extraction. Unlike the planar region, the SAG regions have a lower effective refractive index due to their subwavelength structuring with air gaps and the SiO\textsubscript{2} mask ($n_{SiO\textsubscript{2}}\sim 1.4$)\cite{Kischkat:12}, which could enhance absorption and extraction of light from PbSe. Further, the PbSe-SiO\textsubscript{2} interface acts as a much stronger reflector than the PbSe-GaAs interface, which could result in enhanced absorption or light extraction by reflecting light in the regions where PbSe has overgrown the mask that would otherwise transmit into the substrate in a typical planar film. 

Excitation-dependent PL intensity was used to probe nonradiative recombination pathways on this regrown area and the planar control region; it is plotted as a ratio of the respective PL intensities at each input power in Figure~\ref{fig:PLquality}c. The PL ratio was approximately $\sim$1.0 at the lowest pump power (18.9 mW) with a small positive slope up to 1.2 at $\sim$198.2 mW, where the slope became approximately zero for powers up to 634 mW. Minimal change in PL ratio over two orders of magnitude of pump power indicates that the selective growth process did not have a significant impact on the PbSe material quality and that Auger recombination is similarly dominant at high-injection regimes for the SAG region. Interestingly, the large increase in surface area of the SAG region compared to the planar control did not appear to significantly impact PL signal by adding another nonradiative pathway. This could be due to both the smooth, highly-ordered epitaxial surfaces that form as a result of this growth technique and the decreased sensitivity of PbSe and other IV-VI chalcogenides to surface states. These PL results show that PbSe SAG preserves mid-infrared emission functionality, supporting its use for deterministic integration of nanostructures in hybrid nanowire networks or site-selective quantum emitters. 

\subsection{Gate-tunable Josephson transport in selectively grown PbSe weak links}

Gate-tunable Josephson junction devices are promising candidates for superconducting qubit applications. \cite{larsen_semiconductor-nanowire-based_2015, purkayastha_transmon_2026,sun_junction-intrinsic_2026, luthi_evolution_2018} A non-dissipative supercurrent can be transported between two superconductors with different phases separated by a thin, weak-link material, via Cooper pairs. Group IV-VI semiconductors such as PbTe and PbSe offer potential advantages as building blocks of hybrid superconductor-semiconductor Josephson devices. Their large dielectric constants may help screen disorder and their strong spin-orbit coupling may enable electrically controlled spin states in qubit devices \cite{allgaier_mobility_1958, springholz_wiley_2014,wrasse_quantum_2011,avdeev_valley_2017, huang_disorder_2021,cao_numerical_2022,ten_kate_small_2022,gomanko_spin_2022,jung_selective_2022}. Although PbTe saw progress as a quantum transport platform in recent years, \cite{jung_selective_2022,gupta_evidence_2024,zhang_proximity_2023} PbSe has a comparably interesting yet unexplored potential. 

\begin{figure}
    \centering
    \includegraphics[page=5,scale=.8,clip,trim=0in 3.38in .9in 0in]{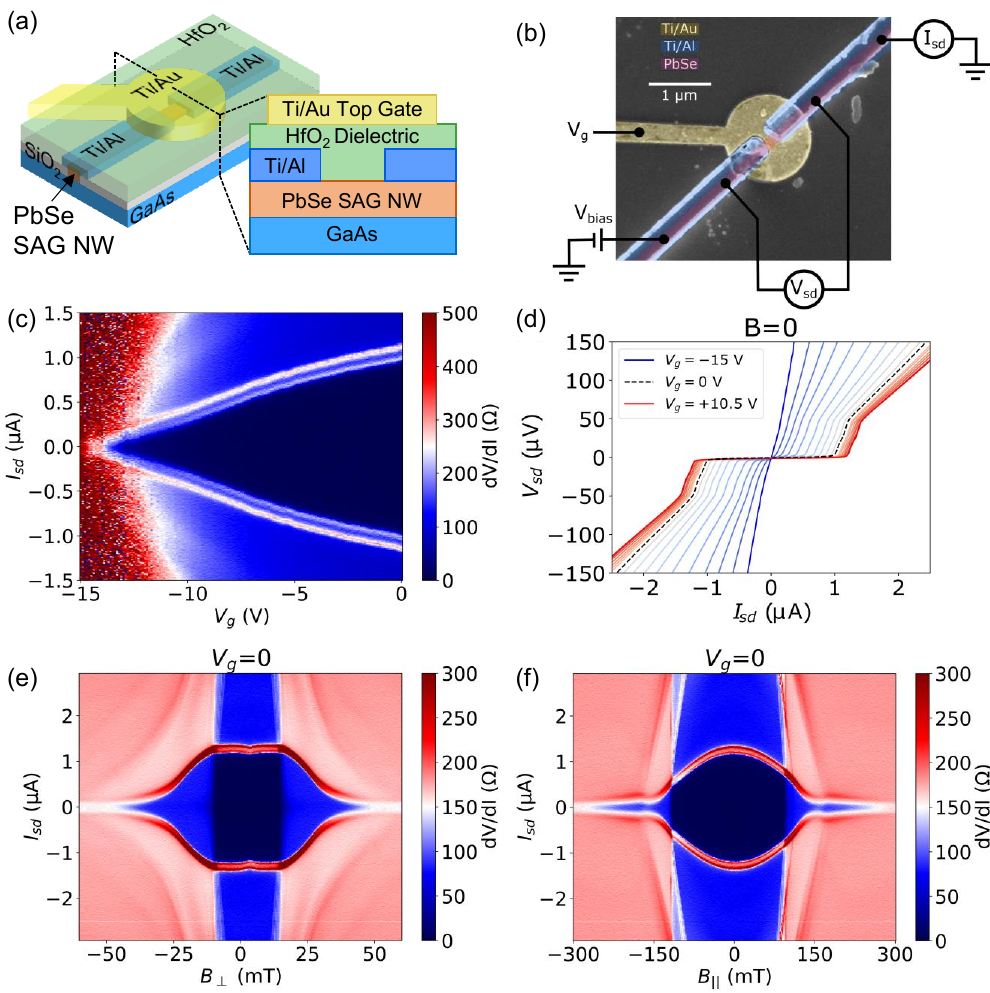}
    \caption{Device schematics and transport measurements for a gate-tunable PbSe nanowire Josephson junction. (a) Cross-sectional diagram showing the device material stack (not to scale). (b) False-color SEM image of the device with the 4-point, voltage-bias measurement schematic. (c) Device I-V characteristics as a function of gate voltage. (d) Device I-V curves with different applied gate voltages ranging from -15V to +10.5V in increments of 1.5V. (e) Differential resistance ($dV/dI$) across the device as a function of current ($I_{SD}$) and out-of-plane magnetic field ($B_{\perp}$). (f) Differential resistance ($dV/dI$) across the device as a function of current ($I_{SD}$) and in-plane field ($B_\parallel$).}
    \label{fig:NWJoFETs}
    
\end{figure}

We fabricated SNS (Superconductor-Normal-Superconductor) Josephson junctions (JJs) with superconducting Al leads contacting SAG PbSe nanowires on an n-type GaAs substrate. A top gate allows for modulation of the carrier concentration in the nanowire weak link to tune the supercurrent across the junction. We demonstrate three key results that lay the groundwork for future quantum transport studies in PbSe-on-GaAs nanowire platforms: we show induced superconductivity in PbSe nanowires coupled to Al superconductor contacts; we show that the Josephson junction critical current can be modulated by means of electrical gating; and finally, we observe superconducting quantum interference under applied perpendicular-to-plane and in-plane magnetic fields.

Figure~\ref{fig:NWJoFETs}a illustrates the stack structure of the JJs. \textcolor{black}{After ion milling, we deposited 5 nm of Ti then 50 nm of Al superconducting contacts using electron beam evaporation. Then 25 nm of HfO\textsubscript{2} dielectric was deposited via atomic layer deposition. Top gates (5 nm Ti/50 nm Au) were then deposited via evaporation.} Six JJs were fabricated and measured on the same chip. Two showed induced superconductivity in the PbSe nanowire, and one of those showed gate tunability as well. \textcolor{black}{See the Methods section for additional details on the JJ fabrication. The results presented in Figure~\ref{fig:NWJoFETs} are all for the electrically gated JJ.} The nanowire dimensions are approximately 150 nm in width, 70 nm in height, and 130 nm in junction length. An SEM image of the measured device is shown in Figure~\ref{fig:NWJoFETs}b along with the experimental schematic for a 4-point, voltage-bias measurement where the current ($I_{SD}$) and voltage ($V_{SD}$) were measured across the junction. A current-bias setup was also used where the current ($I_{SD}$) was sourced and the voltage ($V_{SD}$) was measured across the junction. All measurements were made in a dilution refrigerator at a temperature of $\sim$9 mK. 

Figure~\ref{fig:NWJoFETs}c shows the differential resistance ($dV/dI$) of this JJ as a function of gate voltage value ($V_{g}$) and measured current ($I_{SD}$) in a voltage-bias setup. 
 The dark blue region corresponds to the zero-resistance state of the junction with a tunable critical current, approaching full suppression of the critical current at  $\sim$-15V. The JJ acts like a Josephson Field Effect Transistor (JoFET), which combines FET gate-tunablity with the zero-resistance supercurrent of a JJ, a prerequisite for gate-tunable superconducting qubit applications \cite{larsen_semiconductor-nanowire-based_2015,graziano_transport_2020,deLange2015}. Figure~\ref{fig:NWJoFETs}d shows the I-V curves at different gate voltages in a voltage-bias setup. The gating response indicates n-type electrical transport in the PbSe nanowire. Although the GaAs substrate is also n-type, its contribution to the measured transport is negligible. Parallel current through the substrate beneath the nanowire is suppressed by its substantially higher resistance (on the order of 100 $k\Omega$) relative to the junction’s normal-state resistance \textcolor{black}{($R_{N}\sim$85 $\Omega$ at zero gate voltage).} Although the supercurrent can be suppressed, the normal-state electrical transport can only be modulated by \textcolor{black}{2\%} at an applied gate voltage of -15V. \textcolor{black}{We believe the low $I_{SD}$ modulation under gating is due to the high carrier concentration in the nanowires. Hall measurements of planar PbSe films (80 nm in thickness) revealed carrier concentrations greater than $5\times10^{18}$ cm\textsuperscript{-3} and carrier mobilities below 200 cm\textsuperscript{2}/(Vs).} The estimated mean free path is much shorter than the 130 nm junction length, consistent with transport in the diffusive regime. 

Figures~\ref{fig:NWJoFETs}e and \ref{fig:NWJoFETs}f show the JJ critical current response to an out-of-plane ($B_{\perp}$) and in-plane ($B_{\parallel}$) magnetic field, respectively, in a current-bias setup. The dark blue region bounded by the red corresponds to the junction supercurrent being tuned with the applied magnetic field. The different features in the superconducting behavior of the device correspond to the induced superconductivity in the nanowire and portions of the superconducting contacts (outside the junction) transitioning to the normal state. The resonant features outside the supercurrent region are likely due to the contact and circuit geometry. The superconducting quantum interference pattern features a symmetric decay of the critical current with applied field. The applied field threads the junction area, which creates a position-dependent phase shift across the junction’s width, yielding this spatial interference pattern. Figure~\ref{fig:NWJoFETs}e for the out-of-plane field only features a central lobe, while Figure~\ref{fig:NWJoFETs}f for the in-plane field features a central lobe and two side lobes. The angle between the in-plane B-field and the nanowire is $\sim$35 degrees. These interference patterns imply phase-coherent transport across the junction, as required to establish the Josephson effect. Plots of the out-of-plane and in-plane interference patterns as a function of applied gate voltage are included in the \textcolor{black}{Supplemental Information.} 

\textcolor{black}{The observation of a gate-tunable Josephson supercurrent and a quantum interference pattern in a 130-nm-long Al-PbSe-Al SNS junction demonstrates the superconducting proximity effect in the nanowire weak link. While the present devices operate in the diffusive regime, the electronic quality of the PbSe SAG material is sufficient to support phase-coherent Josephson transport and gate tunability. The PbSe Josephson junction exhibits a critical current density on the order of $10^4$ A/cm\textsuperscript{2}, comparable in magnitude to values reported for PbTe and InAs nanowire JJs, indicating robust superconducting proximity coupling.\cite{gupta_evidence_2024,abay_high_2012} These results provide a promising foundation for further device optimization and for realizing more complex device architectures using selective area growth.}

\section{Conclusion}

In this work, we demonstrated selective area growth and lateral coalescence of highly-mismatched PbSe nanostructures on GaAs. The morphology was highly smooth, controlled and well-ordered, driven by the formation of the low-energy \{001\} facets of PbSe. STEM and ECCI of the partially coalesced islands demonstrated SAG of PbSe as a path for strain and threading dislocation management, separating defect formation in island growth and coalescence. A large fraction of islands as well as coalescence seams remain defect free, despite the large lattice-mismatch. Improvements in the fabrication flow may further reduce dislocation generation. The unique morphology may yield simpler paths towards planar coalescence to achieve complete encapsulation of mask material. PbSe SAG nanostructures show the potential of PbSe selective-area growth for realizing mid-infrared optical structures and quantum devices, with a path toward scalable integration. We observed no degradation in mid-infrared PL from the selectively-grown structures as a result of the technique or increased surface-to-volume ratio at room temperature, and we showed evidence of gate-modulated induced superconductivity and phase-coherent transport from Jo-FETs fabricated on PbSe nanowires. These demonstrations of optical and electrical transport properties motivate PbSe SAG as a promising platform of integration for applications requiring high quantum efficiency and low disorder by combining its intrinsic material properties and etch-free semiconductor patterning.

\section{Methods}
\subsection{Molecular beam epitaxy}
Selective PbSe growth studies were performed over SiO\textsubscript{2} films patterned by e-beam lithography on Si-doped (001) GaAs. The mask openings consisted of arrays of squares with varied widths and periods as well as a pinwheel structure to assess the orientation-dependent faceting during LEO.  A compound source was used to supply a PbSe flux, equivalent to a unity sticking growth rate of 0.4 \AA/s unless otherwise noted and the Se flux was supplied by a valved Se cracker. For all samples in this study, the native oxide of the GaAs wafer was thermally desorbed at $\sim$575-580 $^\circ$C under a high Se flux corresponding to 2$\times10^{-7}$ Torr BEP, and then a surface preparation PbSe dose (30s) was performed at 420 $^\circ$C to promote cube-on-cube nucleation of IV-VI compounds on (001) III-V surfaces as in previous studies.\cite{haidet_nucleation_2020,meyer_bright_2021} Due to the high temperature nature of this step, we observe that the surface pre-treatment does not influence deposition selectivity, which is an encouraging difference in comparison to techniques for single-orientation III-V heteroepitaxial nucleation.\cite{lin_antiphase_2013,huang_simultaneous_2008} that typically require low non-selective temperatures or Al-containing buffers. More details on template fabrication and surface preparation can be found in the Supplemental Information.

\subsection{Structural characterization}
SEM images were taken on a Thermo Fisher Scientific (TFS) Apreo-S microscope at 5 kV. TEM foils were prepared in an FEI Helios NanoLab 600i DualBeam SEM/Focus Ion Beam system. Atomic resolution STEM was then performed using a TFS Spectra 300 with a 300 kV electron beam. 

ECCI was performed using a TFS Apreo S LoVac SEM operated at 30 kV and 1.6 nA with a working distance of 5 mm and dwell time of 100 $\mu$s. The electron channeling pattern was collected in a reference growth region to establish the channeling condition, which was taken at the approximate intersection between the $\{$220$\}$ and $\{$400$\}$ bands. Due to slight height variations between the coalesced and mask areas, two images were taken for each region of interest with the contrast optimized in one image for the laterally overgrown, coalesced areas and the other for the area within the mask, keeping all other imaging conditions constant. Full ECCI images are available in the Supplemental Information.

\subsection{Photoluminescence spectroscopy}
Optical characterization was achieved with quasi-continuous wave PL measurements using an 808 nm laser with an output power of 1 W and a modulation frequency of 10 kHz with a 50\% duty cycle. The laser passed through a 3 $\mu$m dichroic beamsplitter and was focused with an all-reflective objective onto the sample. The PL response from the sample was then collected through the objective, reflected by the dichroic mirror and directed through an anti-reflection-coated (3-5 $\mu$m) Si window for filtering out laser light. An additional 1900 nm longpass filter was incorporated for the excitation-dependent measurements to remove background signal coming from the nGaAs substrate. The PL spectra were captured with a fourier transform infrared (FTIR) spectrometer. 

\subsection{Gate-tunable Josephson junction fabrication}

\textcolor{black}{The 2-terminal Josephson junctions were fabricated using the SAG PbSe nanowires as the semiconductor weak link (deposited via MBE on a GaAs substrate chip). The superconducting contacts were patterned by e-beam lithography using 950K poly(methyl methacrylate) (PMMA) A4 as the resist layer. Before deposition, ion milling of the chip surface was performed with three 10-second bursts separated by 30-second waiting periods. Immediately afterwards, Ti/Al (5 nm/50 nm) were deposited by e-beam evaporation in an ultra-high vacuum system. HfO\textsubscript{2} dielectric was then deposited via atomic layer deposition at 180°C. Finally, the top gates were similarly patterned using the same resist layer as the contacts and Ti/Au (5 nm/50 nm) were deposited by e-beam evaporation. All fabrication was performed in Class 100 cleanrooms. Six JJs were fabricated and measured on the same chip. Two showed induced superconductivity in the PbSe nanowire, and one of those showed gate tunability as well. All the presented results are derived from the electrically gated JJ. The yield was impacted by wire-bonding challenges and leakage through the gate dielectric. All measurements were made in a dilution refrigerator at a temperature of $\sim$9 mK.}

\section*{Author Contributions}
\textbf{Ashlee M. Garc\'ia:} Conceptualization (equal); Data Curation (lead); Investigation (lead); Software (lead); Formal Analysis (lead); Writing - Original Draft (equal); Visualization (lead); Validation (equal); Project Administration (equal).
\textbf{Hosni A. Kaissi:} Conceptualization (supporting); Data Curation (supporting); Investigation (supporting); Formal Analysis (supporting); Writing - Original Draft (supporting); Visualization (supporting); Validation (equal); Project Administration (supporting).
\textbf{Jarod E. Meyer:}
Investigation (supporting); Data Curation (supporting); Formal Analysis (supporting); Writing – Review \& Editing (equal).
\textbf{Kira J. Martin:}
Investigation (supporting); Data Curation (supporting); Formal Analysis (supporting); Writing – Review \& Editing (equal).
\textbf{Maksim Gomanko:} Investigation (supporting); Data Curation (supporting); Formal Analysis (supporting); Writing – Review \& Editing (equal).
\textbf{Laura A. Stern:}
Investigation (supporting); Data Curation (supporting); Formal Analysis (supporting); Writing – Review \& Editing (equal).
\textbf{Pooja D. Reddy:}
Investigation (supporting); Data Curation (supporting); Formal Analysis (supporting); Writing – Review \& Editing (equal).
\textbf{SeongJin Park:}
Investigation (supporting); Data Curation (supporting); Formal Analysis (supporting); Writing – Review \& Editing (equal).
\textbf{Wilson J. Y\'anez-Parre\~no:} Investigation (supporting); Data Curation (supporting); Formal Analysis (supporting); Writing – Review \& Editing (equal).
\textbf{Sergey M. Frolov:} Conceptualization (supporting); Funding Acquisition (equal); Resources (supporting); Supervision (supporting); Writing - Review \& Editing (equal).
\textbf{Vlad S. Pribiag:} Conceptualization (supporting); Funding Acquisition (equal); Resources (supporting); Supervision (supporting); Validation (supporting); Writing - Review \& Editing (equal).
\textbf{Kunal Mukherjee:} Conceptualization (equal); Funding Acquisition (equal); Resources (lead); Supervision (supporting); Validation (supporting); Writing - Original Draft (equal); Project Administration (equal).

\section*{Acknowledgments}
Molecular beam epitaxy studies, template fabrication and electrical transport studies were supported by the US Department of Energy (DE‐SC0019274). Surface preparation prior to epitaxial growth was done in part at nano@stanford RRID:SCR$\_$026695. We gratefully acknowledge support for PL spectroscopy and transmission electron microscopy via the UC Santa Barbara NSF Quantum Foundry funded via the Q-AMASE-i program under award DMR-1906325. K. M. acknowledges William E. McMahon for helpful discussions. J. E. M. gratefully acknowledges support from the TomKat Center for Sustainable Energy's TomKat Center Graduate Fellow for Translational Research Fellowship. Josephson junction nanofabrication work for this project was conducted in the Minnesota Nano Center (MNC). The cost of MNC user fees for this nanofabrication work was supported by the DOE under award No. DESC0019274. The MNC received partial support for its operations from the National Science Foundation through the National Nano Coordinated Infrastructure Network (NNCI) under Award Number ECCS-2025124.

\section*{Conflicts of Interest}
V. S. P. is founder and CEO of D-Rez LLC, which leverages superconductor-semiconductor hybrid materials to develop computing hardware.
\section*{Data Availability Statement}
Data underlying the results presented in this paper are not publicly available at this time but may be obtained from the authors upon reasonable request.


\bibliography{achemso-demo}
\newpage

\noindent\textbf{\Large Supplemental Information}

\setcounter{figure}{0}
\makeatletter
\renewcommand{\thefigure}{S\@arabic\c@figure}
\makeatother

\begin{figure}
    \centering
    \includegraphics[page=6,scale=0.8,clip,trim=0in 7.69in 3.54in 0in]{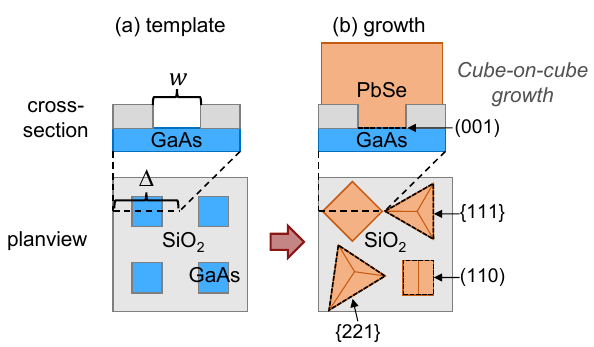}
    \caption{(Top) Cross-section and (bottom) planview schematics showing (a) fabricated template and (b) orientation of nuclei post-growth. Cube-on-cube growth is the ideal mode, where \{001\} PbSe nucleates on \{001\} GaAs, however presence of pyramidal and tent-like growth indicate misnucleation of the \{111\}/\{221\} and \{110\} planes, respectively, on the \{001\} GaAs surface.}
    \label{fig:misnucleation}
    
\end{figure}

\noindent\textbf{Substrate fabrication}

 Patterned \ce{SiO2} on (001) nGaAs templates were fabricated at the University of Pittsburgh. Initial fabrication process used an e-beam lithography with MMA/PMMA resist stack (100kV 250/20nA beams, 1000 $\mu$C/cm\textsuperscript{2} dose) on GaAs to pattern alignment marks; the wafers are then developed in isopropanol:deionized (DI) wafer for 12s at room temperature (or 2 min 6 $^\circ$C). The exposed areas are then etch in an inductively coupled plasma reactive ion etcher (ICP-RIE) with a gas chemistry of 7 sccm \ce{Cl2}:20 sccm \ce{BCl3}, 75 W bias and ICP of 600W at a chamber pressure of 3 mTorr. Resist is then removed with acetone and isopropanol. The second round of fabricated templates in this study, referred to as “optimized templates” or "Fab $\#$2", do not undergo this initial lithography step and instead start with the following deposition process. The \ce{SiO2} film is deposited using plasma-enhanced chemical vapor deposition (PECVD); the film thicknesses in this study range from $\sim$47-83 nm. ZEP520a resist is spun on the wafer, exposed in 100kV EBPG ($\sim$230 $\mu $C/cm\textsuperscript{2} dose, 250/20/4 nA beams for different feature sizes), developed in amyl acetate 1:50s at room temperature and rinsed in isopropanol for 30s. The \ce{SiO2} film is etched with a 30:10 sccm \ce{CF3}:\ce{CHF4} gas chemistry in the ICP-RIE (50W bias/600W ICP at 10mTorr), which has an approximate etch rate of $\sim$200 nm/min. Remove remaining resist with n,n dimethyl acetamine and then a 30s \ce{O2} plasma treatment. Finally, the templates are cleaned in acetone and isopropanol. 

\vspace{2em}

\noindent\textbf{Surface preparation}

MBE growth of PbSe was performed on a Riber Compact 21 system, equipped with a PbSe compound effusion cell and a valved cracker to supply the Se flux. Prior to loading fabricated templates for growth, the top epi-surface is oxidized in a Samco UV-Ozone Cleaner at 150 $^\circ$C for 5 min. After which, the oxide is removed with a 1:10 HCl rinse for 90 seconds and rinsed in water for 2 min. The Samco and acid treatments were performed in the Stanford Nano Shared Facilities. Both the \ce{SiO2}/nGaAs template and a semi-insulating (SI) GaAs 1 cm$^2$ control undergo this treatment and then are indium bonded to a platen such that mask-free wafer can be used for temperature measurement and reflection high energy electron diffraction (RHEED) to observe crystal growth. The platens are then loaded into the load chamber and a bake is performed. The cassette of platens is transferred into the buffer chamber and the Ti-sublimation pump is run at the beginning of the growth day. 

\begin{figure}
    \centering
    \includegraphics[page=7,scale=0.8,clip,trim=0in 7.37in 1.81in 0in]{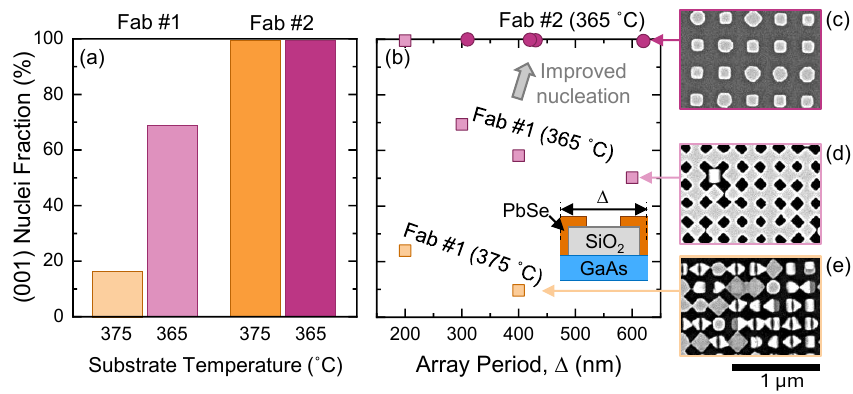}
    \caption{(a) Bar graph comparing the fraction of cube-on-cube-orientated nucleation sites for Fab\#1 and Fab\#2 templates at different substrate temperatures. (b) Fraction of (001) nucleation sites with respect to array period, $\Delta$, for both fabrication runs. Representative planview SEM images of nucleation from (c) Fab \#2 at 365$^\circ C$, (d) Fab \#1 at 365$^\circ C$ and (e) Fab \#1 at 375$^\circ C$.}
    \label{fig:temp-and-template}
    
\end{figure}

Prior to growth, each platen undergoes a 50 min buffer chamber bake and then is transferred into the growth chamber. The first step in the growth process is thermal desorption of the native oxide under a high Se overpressure. The wafers are brought to a temperature at which thermal desorption is observed in RHEED, typically $\sim$575-580 $^\circ$C, measured by optical pyrometry. The substrate is held for 10 min to ensure complete removal of native oxide and the Se valve is closed once the substrate temperature reaches 500 $^\circ$C. After which, the dose step and subsequent heteroepitaxial growth can be performed.

\vspace{2em}

\begin{figure}[h]
    \centering
    \includegraphics[page=8,scale=0.8,clip,trim=0in 7.62in .81in 0in]{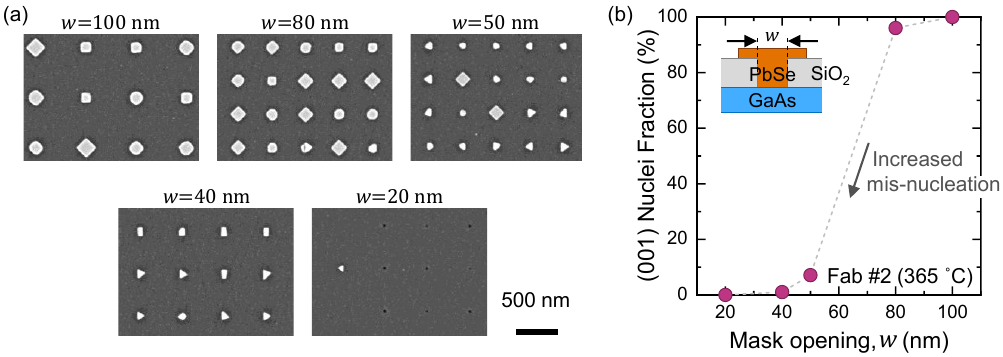}
    \caption{(a) Representative SEM images of nucleation within $w\leq$100 nm square openings showing presence of mis-nucleation as the mask opening decreases. (b) Graph of fraction of cube-on-cube nucleation with respect to square mask opening width, $w$.}
    \label{fig:mask-opening}
    
\end{figure}

\noindent\textbf{Controlling misnucleation}

It was observed that variation in the growth template and mask fabrication had significant influence on the control of the nucleation. A schematic comparing the planview shape of of misoriented nuclei in comparison to cube-on-cube growth is included in Figure~\ref{fig:misnucleation} to aid the clarity of the discussion. As shown in Figure~\ref{fig:temp-and-template}, growth at 367 and 375 $^\circ$C achieved near-complete cube-on-cube nucleation for Fab $\#$2 templates, while increasing substrate temperature and spacing between mask opening both resulted in greater rates of mis-nucleation (i.e. the formation of (110)-, (221)-, or (111)-oriented nuclei) for Fab $\#$1 templates. The origin of the mis-nucleation is hypothesized to be a result of either contamination of the initial episurface or damage from over-etching of the oxide. Interestingly, there was evidence of mis-nucleation on Fab $\#$2 templates as the mask opening decreases (Figure~\ref{fig:mask-opening}) to 40 and 50 nm wide, which could be a result of incomplete etching of these narrow openings or the presence of nucleation sites from etch-induced roughness on the \ce{SiO2} sidewalls[1] that become more concentrated and therefore favorable as the opening becomes narrow.

\vspace{2em}

\noindent\textbf{Deposition selectivity}

\begin{figure}
    \centering
    \includegraphics[page=9,scale=0.8,clip,trim=0in 7in 3.5in 0in]{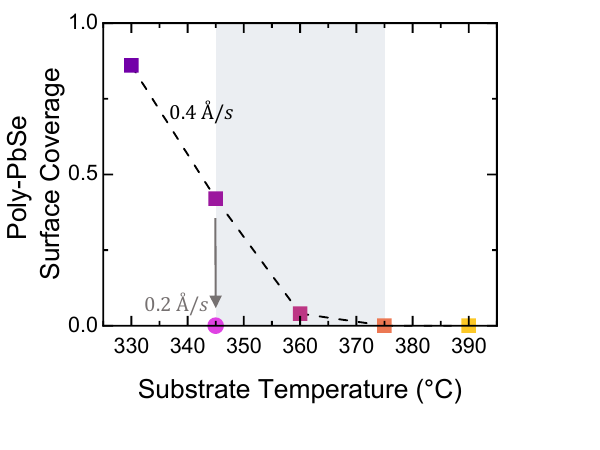}
    \caption{Graph of the area fraction of polycrystalline PbSe deposition covering the \ce{SiO2} mask (i.e., surface coverage) as a function of substrate temperature for two different growth rates.}
    \label{fig:mask-opening-dep}
    
\end{figure}

 PbSe deposition selectivity with respect to growth temperature is represented by polycrystalline surface coverage fraction for PbSe deposition flux corresponding to a growth rate of 0.4 \AA/s. By reducing the growth rate to 0.2 \AA/s, selective growth can be achieved at 345 $^\circ$C, enabling compatibility with PbSnSe[2-4] and PbGeSe[5] growth regimes (Figure~\ref{fig:mask-opening-dep}).

\noindent\textbf{Nanowire morphology}

Continuous 100 nm wide nanowires were observed under optimized 365$^\circ C$ growth conditions. Any instances of misoriented nucleation that is observed in the island growth was not presence likely due to the coalescence of cube-on-cube islands burying the misoriented grains. Similar to the PbSe islands, the nanowire morphology was dominated by \{001\}-oriented facets due to the energetic favorability of this crystal plane.
\begin{figure}[H]
    \centering
    \includegraphics[page=13,scale=1,clip,trim=0in 5.54in 4.10in 0in]{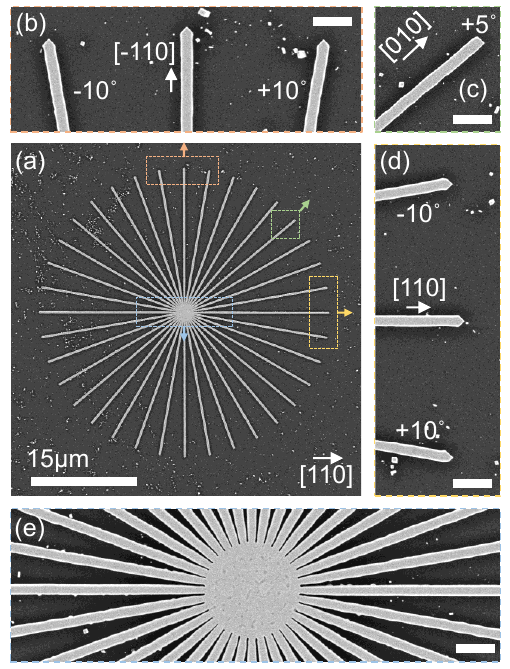}
    \caption{(a) Planview SEM image show growth morphology of pinwheel structure consisting of 100nm wide openings after 180 nm of PbSe growth at 365 $^\circ$C over a 50 nm tall \ce{SiO2} mask. Zoomed in SEM images showing the characteristic \{001\}-driven morphology of the wires along the (b) [-110](0$^\circ$, $\pm$10$^\circ$), (c) [010]($+$5$^\circ$) and (d) [110](0$^\circ$, $\pm$10$^\circ$) directions as well as the center of the pinwheel comparing the rough morphology of large planar regions to that of the visually more uniform wires. (b)-(e) Inset scale bars are 1 $\mu$m.}
    \label{fig:pinwheel}
    
\end{figure}

\vspace{2em}

\noindent\textbf{Electron channeling contrast imaging (ECCI)}

%Below is a table of measured threading dislocation (TD) statistics and the associated ECCI images discussed in the article:
Below are the ECCI images discussed in the article followed by a table of the measured threading dislocation (TD) statistics:
% Requires: \usepackage{booktabs}
\begin{figure}[H]
    \centering
    \includegraphics[scale=1,clip,trim=0in 2.5in 3.75in 0in]{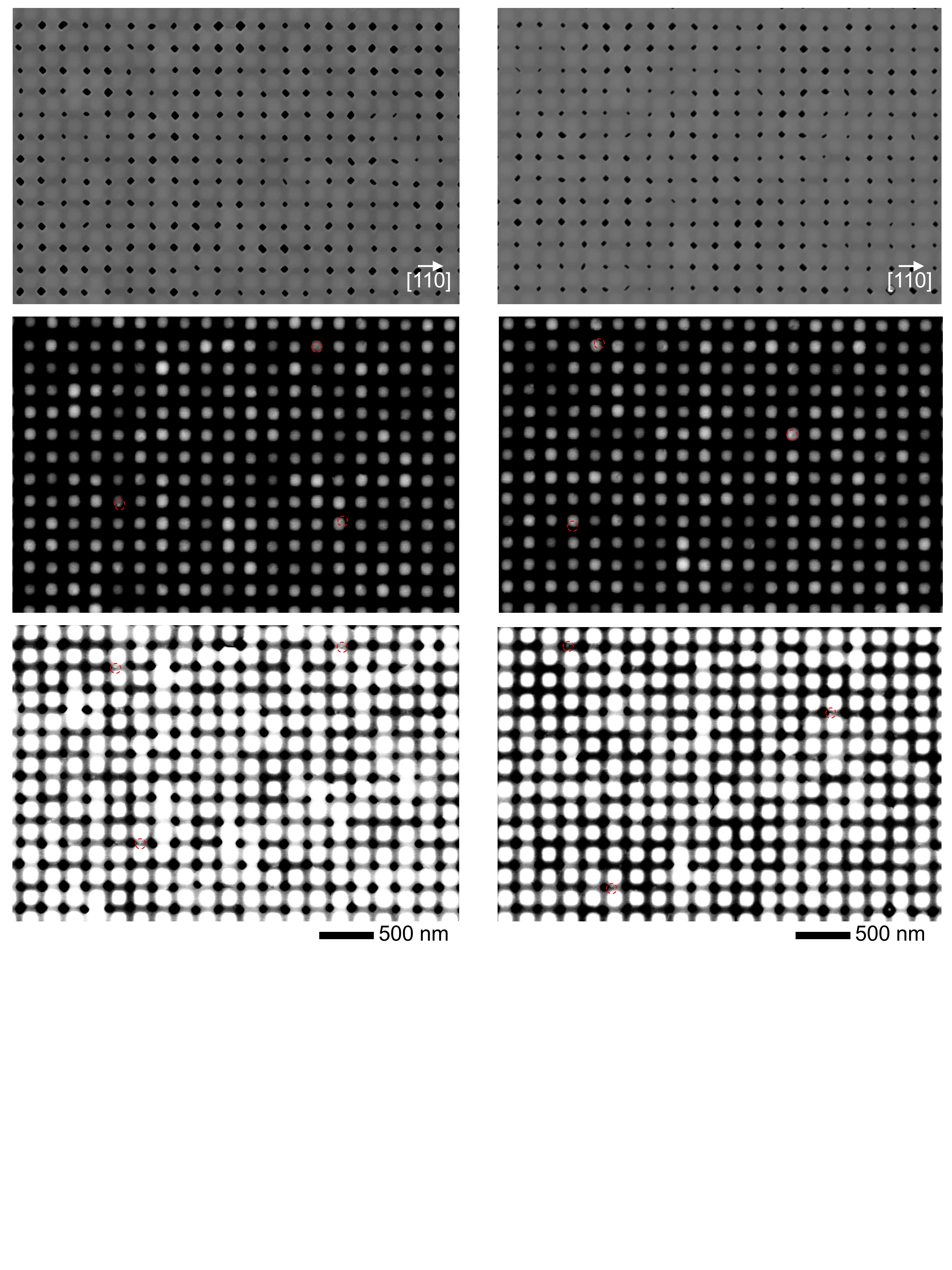}
    \caption{(Top) SEM image of R1 and corresponding ECCI images for (middle) growth within the \ce{SiO2} and (bottom) lateral overgrown regions containing coalesced interfaces. Red dashed circles highlight three threading dislocations in each measurement.}
    \label{fig:ecciR1}
    
\end{figure}

\begin{figure}[H]
    \centering
    \includegraphics[scale=1,clip,trim=3.75in 2.5in 0in 0in]{Slide10.PNG}
    \caption{(Top) SEM image of R2 and corresponding ECCI images for (middle) growth within the \ce{SiO2} and (bottom) lateral overgrown regions containing coalesced interfaces. Red dashed circles highlight three threading dislocations in each measurement.}
    \label{fig:ecciR2}
    
\end{figure}

\begin{figure}[H]
    \centering
    \includegraphics[width=1\linewidth,clip,trim=0in 5in 0in 0in]{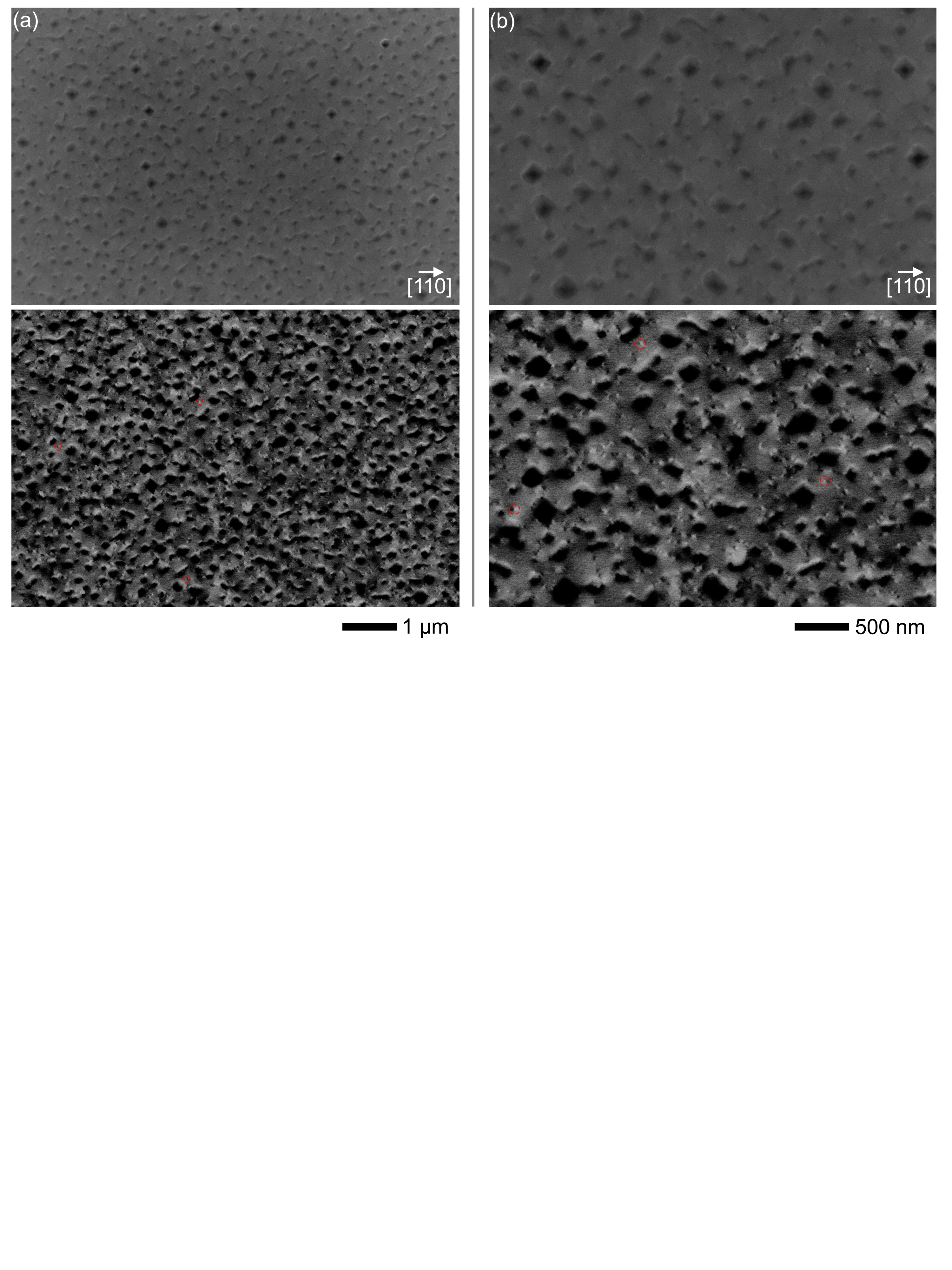}
    \caption{(Top) SEM image and (bottom) corresponding ECCI measurement for planar control region at (a) lower and (b) higher magnification. R3 TD counting corresponds to part (b) due to high density of threading dislocations. Red dashed circles highlight three threading dislocations in each measurement.}
    \label{fig:ecciplanar}
    
\end{figure}

\begin{figure}[H]
    \centering
    \includegraphics[width=1\linewidth,clip,trim=0in 6in 0in 0in]{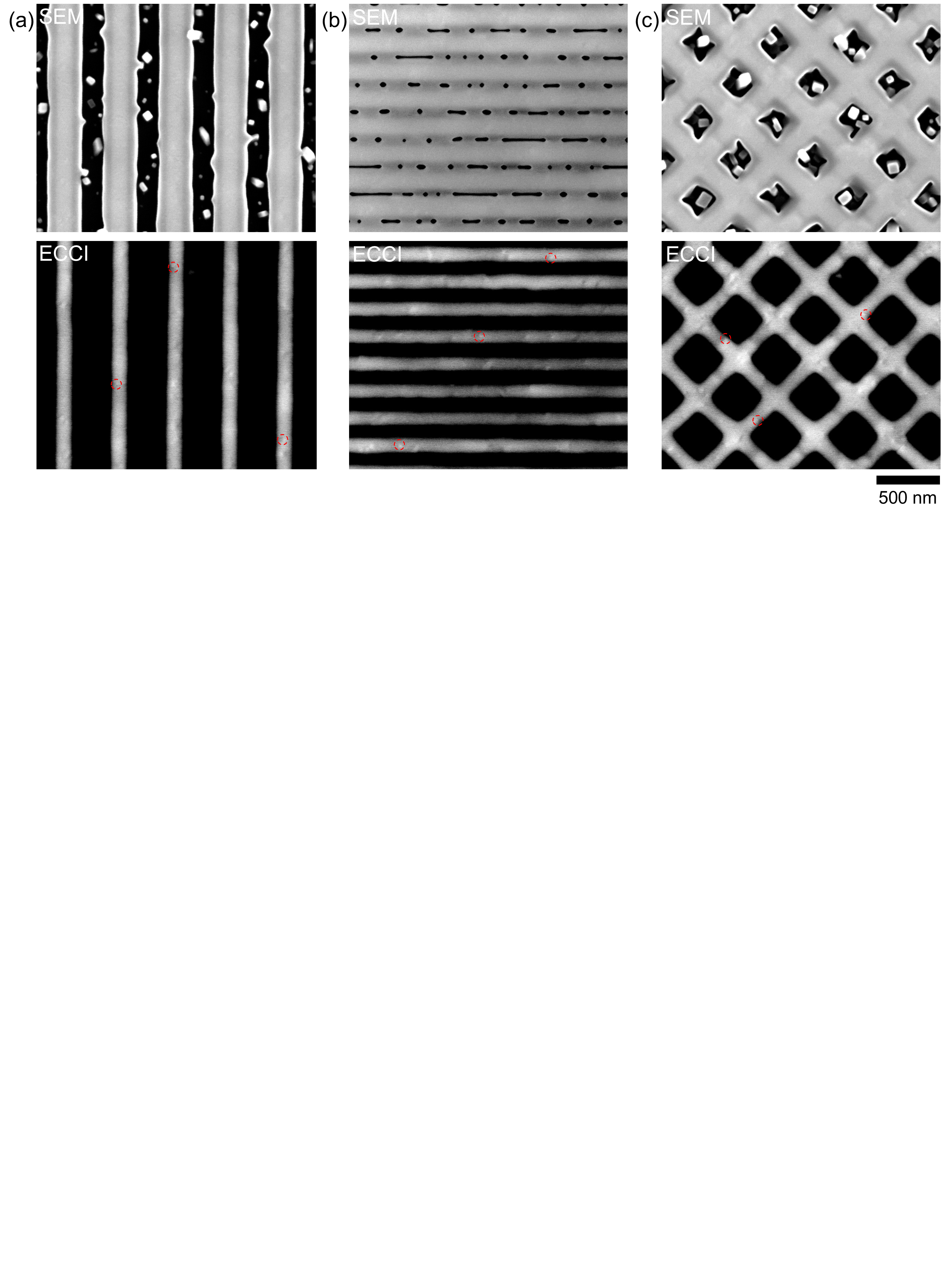}
    \caption{(Top) SEM image and (bottom) corresponding ECCI measurement for 100 nm wide nanowires aligned to the (a) [$\Bar{1}$10], (b) [110] and (c) $<$010$>$ directions. Red dashed circles highlight three threading dislocations in each measurement.}
    \label{fig:ecciwires}
    
\end{figure}

\begin{table}[h]
    \centering
    \caption{Table containing measured areas and TD counts for the associated ECCI measurements below. The array of square openings in an \ce{SiO2} mask is defined by the opening width $w$ and the array period $\Delta$.}
    \label{tab:placeholder_label}
    \begin{tabular}{c|c|c|c|c|c|c}
        \toprule
        \textbf{Meas.} & \textbf{Mask} & \textbf{Region} & \textbf{Area} & \textbf{TD} & \textbf{TDD} & \textbf{TD-free} \\
        \textbf{ID} & \textbf{Pattern} & \textbf{Type} & \textbf{($\mu m^2$)} & \textbf{count} & \textbf{($cm^{-2}$)} & \textbf{Fraction} \\
        \midrule
        R1 & $w$=100 nm & Island & 3.329 & 96 & 2.88$\times 10^9$ & 71.5$\%$ \\
         & ($\Delta$=205 nm) & Merge & 5.618 & 117 & 2.08$\times 10^9$ & 83$\%$\\
         & & \textit{Average} & 8.947 & 213 & 2.38$\times 10^9$ & --\\
        \midrule
        R2 & $w$=100 nm & Island & 3.458 & 73 & 2.11$\times 10^9$ & 76.9$\%$\\
         & ($\Delta$=205 nm) & Merge & 5.888 & 122  & 2.07$\times 10^9$ & 88.7$\%$\\
         & & \textit{Average} & 9.346 & 195 & 2.08$\times 10^9$ & --\\
         \midrule
        R3 & none, planar film & -- & 9.41 & 647 & 6.88$\times 10^9$ & -- \\
        \bottomrule
    \end{tabular}
\end{table}

\noindent\textbf{PbSe island tilt}

Scanning moir\'e fringes were generated by setting the STEM scan sampling close to the PbSe (002) lattice-plane spacing of approximately 0.306 nm. The interference between the scan raster and the crystal lattice produces a longer-period moir\'e fringe pattern. The ratio of the moir\'e and lattice spacings gives the moir\'e magnification, M = d$_{moire}$/d$_{002}$. For small lattice rotations, the same factor relates the moir\'e-fringe tilt to the underlying lattice tilt. The measured moire period was approximately 4.4 nm, corresponding to M approximately 14.4.

\begin{figure}[H]
    \centering
    \includegraphics[page=14,width=1\linewidth,clip,trim=0in .7in 0.2in 5.0in]{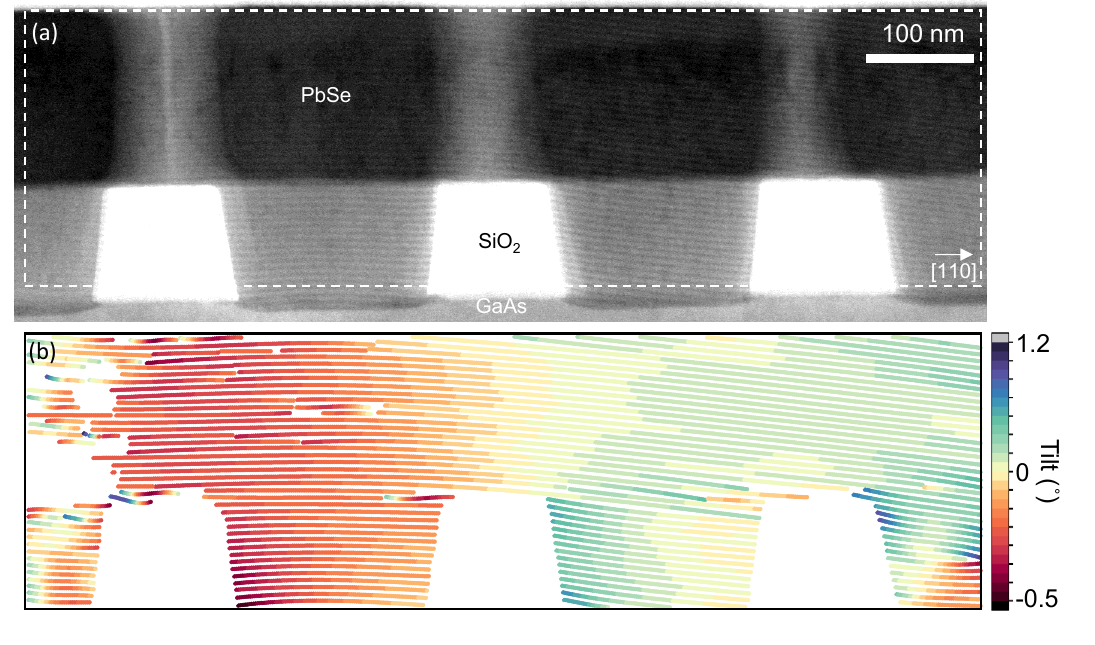}
    \caption{(a) Cross-sectional bright field scanning transmission electron microscopy image showing a moir\'e fringe pattern and (b) corresponding colormap of measured tilt of PbSe planes calculated by taking into account the moir\'e magnification.}
    \label{fig:moire_2}
    
\end{figure}

\noindent\textbf{\textcolor{black}{Quantum interference patterns}}

\begin{figure}[H]
    \centering
    \includegraphics[page=15,width=0.9\linewidth,clip,trim=0in 2.2in 0in 0in]{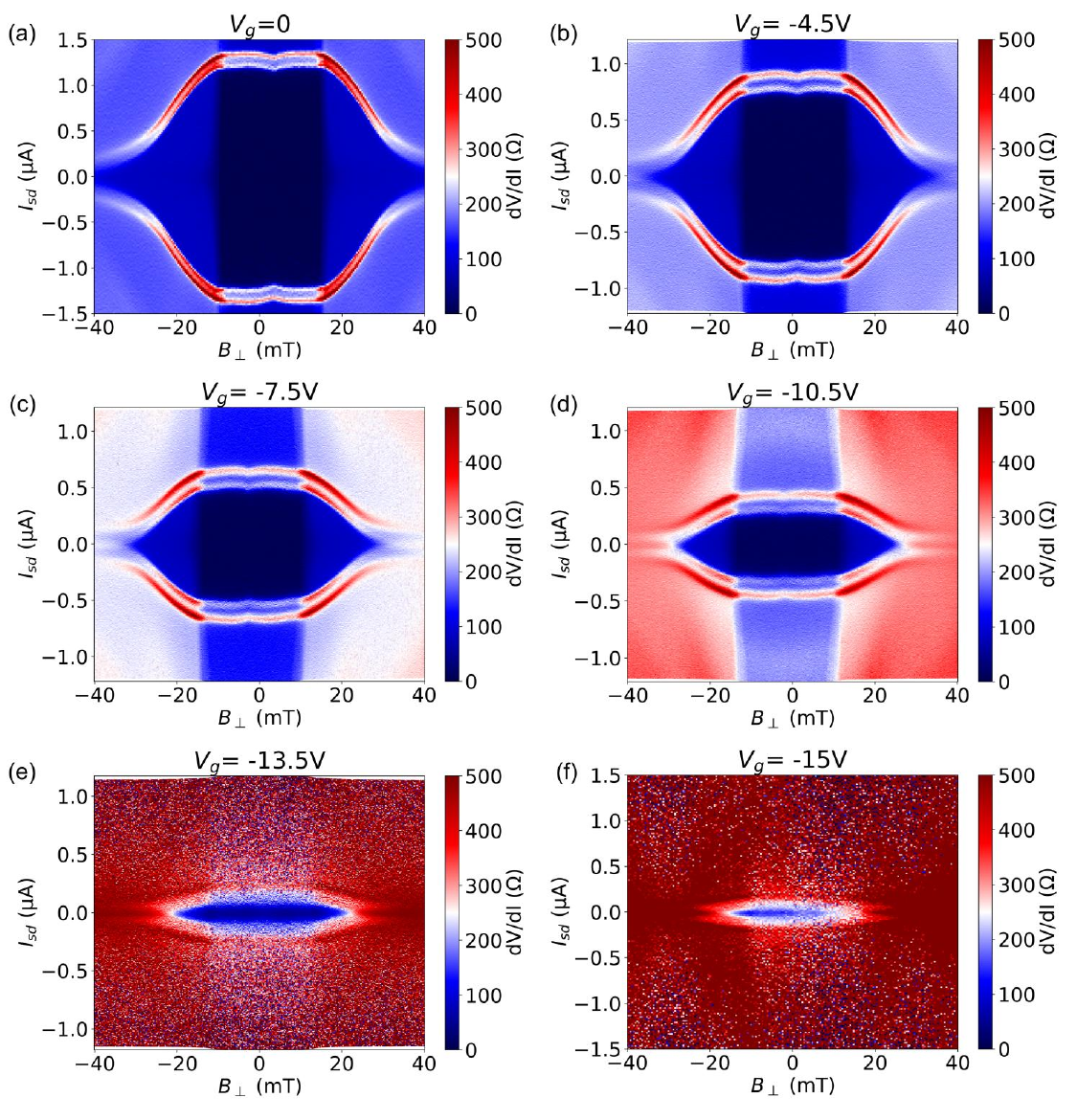}
    \caption{\textcolor{black}{Additional transport measurements for the 2-terminal Josephson junction device. (a-f) Plots for different gate voltages of differential resistance ($dV/dI$) as a function of current ($I_{SD}$) and out-of-plane magnetic field ($B_\perp$).}}
    \label{fig:quantum1}
    
\end{figure}

\begin{figure}[H]
    \centering
    \includegraphics[page=16,width=0.9\linewidth,clip,trim=0in 2.2in 0in 0in]{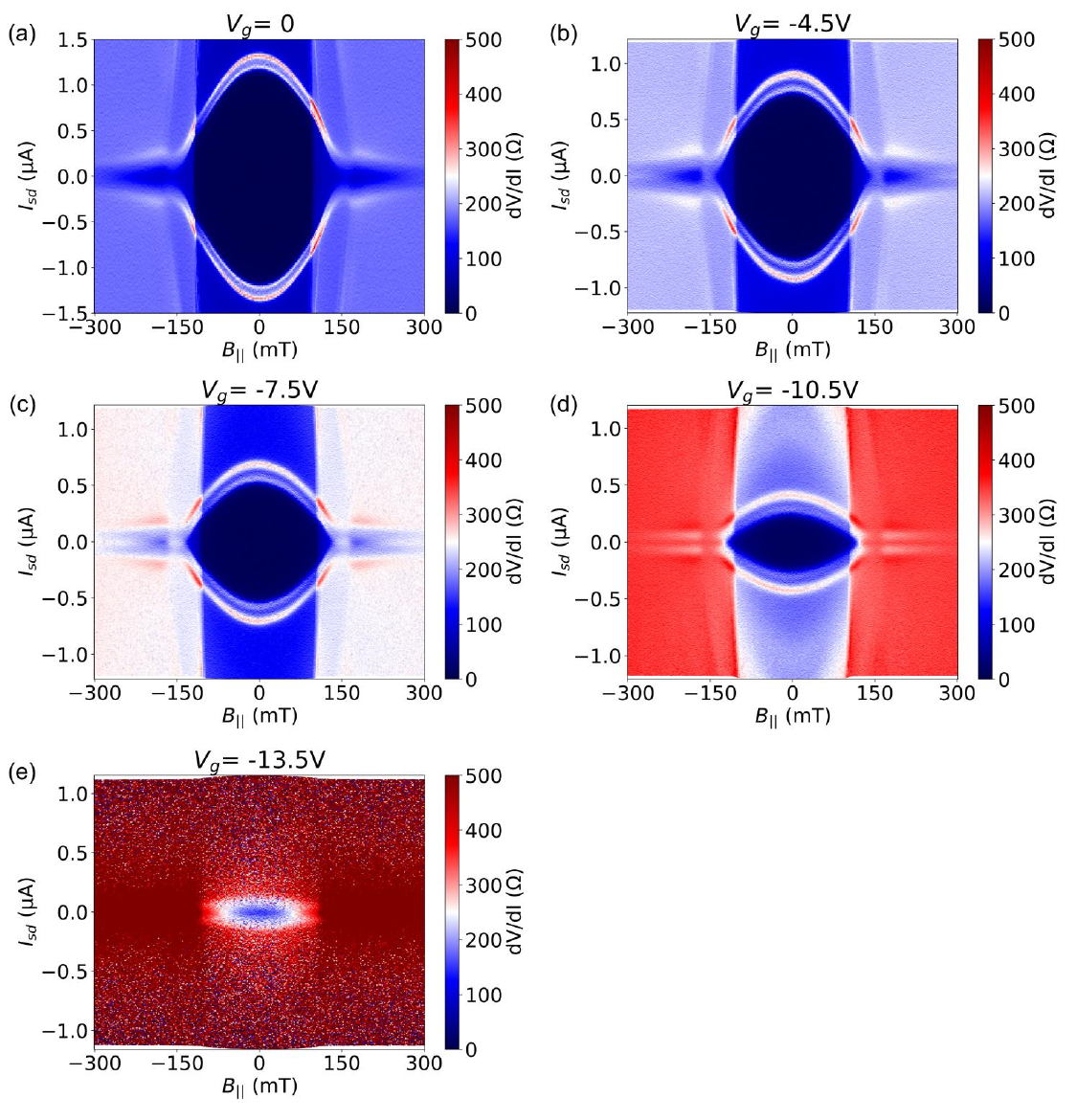}
    \caption{\textcolor{black}{Additional transport measurements for the 2-terminal Josephson junction device. (a-f) Plots for different gate voltages of differential resistance ($dV/dI$) as a function of current ($I_{SD}$) and in-plane magnetic field ($B_\parallel$).}}
    \label{fig:quantum2}
    
\end{figure}

\textcolor{black}{Supplementary Figure~\ref{fig:quantum1} and Figure~\ref{fig:quantum2} plot the interference pattern for out-of-plane and in-plane B-fields as a function of applied gate voltage, respectively. The supercurrent region shrinks with increasing negative gate voltage as the superconducting transport channels are depleted. The measurements were made using the voltage-bias setup from Figure~\ref{fig:NWJoFETs}b.}

\begin{enumerate}[label={(\arabic*)}]
    \item A. M. Garc\'ia, B. D. Aguilar, W. J. Doyle, P. U. Fathi, F. Capasso, D. Wasserman, and S. R. Bank, “Surface Modification for III-V Selective Area Molecular Beam Epitaxy of Non-Selective Mask Materials,” 2026. arXiv eprint: 2606.02317.
    \item J. Meyer, L. Nordin, R. A. Carrasco, P. T. Webster, M. Dumont, and K. Mukherjee, “Engineering PbSnSe Heterostructures for Luminescence Out to 8 $\mu$m at Room Temperature,” Advanced Optical Materials, 2024.
    \item P. Reddy, L. Nordin, L. Hughes, A.-K. Preidl, and K. Mukherjee, “Expanded Stability of Layered SnSe-PbSe Alloys and Evidence of Displacive Phase Transformation from Rocksalt in Heteroepitaxial Thin Films.,” ACS nano, 2024.
    \item P. Reddy, V. Tara, A. Vailionis, A. Majumdar, and K. Mukherjee, “Reversible polymorph switching in IV-VI thin films with epitaxial control and birefringence contrast,” Nano Letters, 2025.
    \item K. Xiao, B. Wong, J. Meyer, and L. Nordin, “Epitaxial PbGeSe thin films and their photoluminescence in the mid-wave infrared,” Journal of Applied Physics, 2024.
\end{enumerate}

\end{document}